\documentclass[manuscript]{aastex}

\shorttitle{Expanded iron UTA spectra as AGN cloud stability probes}
\usepackage{natbib}
\usepackage{index}
\usepackage{color}
\usepackage{aas_macros}
\font\manual=manfnt at 7pt \def\dbend{\hbox{\raise0.9ex\hbox{\manual\char127\hspace{0.6em}}}}

\newcommand\Ion[2]{\ensuremath{\mathrm{#1\,\scriptstyle #2}}}

\newcounter{INTERNALionstage}
\providecommand{\ion}[2]{% replace the aastex version
  \setcounter{INTERNALionstage}{#2}%
  \Ion{#1}{\Roman{INTERNALionstage}}}

\def\gtsim{\mathrel{\hbox{\rlap{\hbox{\lower4pt\hbox{$\sim$}}}\hbox{$>$}}}}
\def\lesssim{\mathrel{\hbox{\rlap{\hbox{\lower4pt\hbox{$\sim$}}}\hbox{$<$}}}}

\def\ps{{\rm\thinspace s^{-1}}}
\def\pcc{{\rm\thinspace cm^{-3}}}

\DeclareMathAlphabet{\vib}{OML}{cmm}{m}{it}
\begin{document}


\title{Expanded Iron UTA spectra -- probing the thermal stability limits 
in AGN clouds}

\author{
G. J. Ferland\altaffilmark{1},
R. Kisielius\altaffilmark{2}, 
F. P. Keenan\altaffilmark{3}, 
P. A. M. van Hoof\altaffilmark{4},
V. Jonauskas\altaffilmark{2},
M. L. Lykins\altaffilmark{1},
R. L. Porter\altaffilmark{5},
R. J. R. Williams\altaffilmark{6}
}

\altaffiltext{1}{Department of Physics and Astronomy, University of Kentucky, 
                 Lexington, KY 40506, USA \email{gary@pa.uky.edu}}
\altaffiltext{2}{Institute of Theoretical Physics and Astronomy, Vilnius 
                 University, A. Go{\v s}tauto 12, LT-01108, Lithuania}
\altaffiltext{3}{Astrophysics Research Centre, School of Mathematics and Physics, Queen's University 
                 Belfast,    Belfast BT7 1NN, UK}
\altaffiltext{4}{Royal Observatory of Belgium, Ringlaan 3, 1180 Brussel, 
                 Belgium}
\altaffiltext{5}{Department of Physics, University of Georgia, Athens, 
                 Ga 30602, USA}
\altaffiltext{6}{AWE plc, Aldermaston, Reading RG7 4PR, UK}

%\author{G. J. Ferland}
%\affil{Department of Physics, University of Kentucky,
%   Lexington, Ky 40506, USA}
%\email{gary@pa.uky.edu}

%\author{R. Kisielius}
%\affil{Institute of Theoretical Physics and Astronomy, Vilnius University,
%A. Go{\v s}tauto 12, Vilnius LT-01108, Lithuania}

%\author{F. P. Keenan}
%\affil{Department of Physics and Astronomy, Queen's University of Belfast,
%   Belfast BT7 1NN, UK}

%\author{P.A.M. van Hoof}
%\affil{Royal Observatory of Belgium, Ringlaan 3, 1180 Brussel, Belgium}

%\author{V. Jonauskas}
%\affil{Institute of Theoretical Physics and Astronomy, Vilnius University,
%A. Go{\v s}tauto 12, Vilnius LT-01108, Lithuania}

%\author{M. L. Lykins}
%\affil{Department of Physics, University of Kentucky, Lexington, Ky 40506, USA}

%\author{R. L. Porter}
%\affil{Department of Physics, University of Georgia, Athens, Ga 30602, USA}

%\author{R. J. R. Williams}
%\affil{AWE plc, Aldermaston, Reading RG7 4PR, UK}

\begin{abstract}
The Fe unresolved transition array (UTAs)  produce
prominent features in the $\sim 15 - 17$ \AA\ wavelength range
in the spectra of Active Galactic Nuclei (AGN). 
Here we present new calculations of the energies
and oscillator strengths of inner-shell lines from Fe\,{\sc xiv}, Fe\,{\sc xv}, 
and Fe\,{\sc xvi}. These are crucial ions since they are dominant at inflection 
points in the gas thermal stability curve, and UTA excitation followed by
autoionization is an important ionization mechanism for these species. We 
incorporate these, and data reported in previous papers, into the plasma 
simulation code Cloudy. 
This updated physics is subsequently employed to reconsider the thermally 
stable phases in absorbing media in Active Galactic Nuclei.
We show how the absorption profile of the Fe\,{\sc xiv} UTA depends on density,
due to the changing populations of levels within the ground configuration.
\end{abstract}

\section{Introduction}
\label{intro}

X-ray observations of active galactic nuclei (AGN) have revealed the existence
of a broad absorption feature in the 15--17 \AA\ wavelength range. This feature
arises from an unresolved transition array (UTA) due primarily to
2$p$ $\rightarrow$ 3$d$ inner-shell absorption lines in iron ions with
an open $M$ shell (Fe\,{\sc i} -- Fe\,{\sc xvi}), and was first identified by
\citet{Sako2001} in the {\em XMM-Newton} spectrum of the quasar IRAS
13349+2438. Since then, the UTA has been detected in numerous AGN (see, for
example, \citet{Longinotti2010}, \citet{Lestinsky2009} and references therein).
It is believed to arise in the warm, relatively low ionization, absorbing gas
surrounding the central supermassive black hole in the AGN.

\citet{Behar2001} first noted the diagnostic potential of the UTA, and showed
that their central wavelengths, absorption profiles and equivalent widths can 
provide information on the ionization structure, kinematics and column densities
of the warm absorbing material in the AGN. These authors also produced atomic 
data required for the modeling of the UTA, including energy levels (wavelengths)
and oscillator strengths for the 16 iron charge states Fe\,{\sc i} through
Fe\,{\sc xvi}, calculated with the multiconfiguration, relativistic {\sc hullac}
computer code \citep{hullac}. Since then, several authors have calculated
wavelengths and oscillator strengths for the UTA. These include \citet{fe1514},
who produced atomic data for Fe\,{\sc xv} and Fe\,{\sc xvi} using the
configuration interaction code {\sc civ3} \citep{civ3} with the inclusion of
relativistic effects by adding Breit-Pauli operators to the
Hamiltonian \citep{breitp}. More recently, \citet{Gu2006} calculated  results
for Fe\,{\sc vi} through Fe\,{\sc xvi} employing second-order many-body
perturbation theory \citep{Lindgren74}. Furthermore, \citet{Beiersdorfer2012}
used the relativistic multi-reference M{\o}ller-Plesset perturbation theory to 
calculate the energy levels of Fe\,{\sc xvi}, including those of the 
autoionizing levels with a hole state in the L shell. They obtained very good 
agreement between the calculated L-shell transition wavelengths and those from 
recent laboratory measurements.

Here we present extended calculations of UTA energies and radiative transition 
rates for Fe\,{\sc xiv}, Fe\,{\sc xv} and Fe\,{\sc xvi}. These are combined 
with other improvements in the plasma simulation code Cloudy  \citep{Ferland13} 
to generate a range of models of absorbing clouds in AGN.
Calculations were performed with r6433 of the Cloudy development branch,
and will be part of the 2013 release of the code.
We also perform a thermal stability analysis to identify the thermally stable 
phases where clouds can exist, and show that Fe\,{\sc xiv}, Fe\,{\sc xv} and 
Fe\,{\sc xvi} probe the high ionization end of the warm stable phase.

\section{New atomic data for Fe\,{\sc xiv}, Fe\,{\sc xv} and Fe\,{\sc xvi} 
transitions} 
\label{fedata}

Previously, a comparatively small set of inner-shell photoexcitation data for
Fe~{\sc xv} and Fe~{\sc xvi} was reported by \citet{fe1514}. In that work energy
levels were identified, and electric dipole transition wavelengths, oscillator 
strengths and transition probabilities determined. Specifically, transitions 
involving photoexcitation from the inner 2$p$ shell to the outer $n=3$ shell 
were investigated using the relativistic Breit-Pauli approach, and an assessment
of the accuracy of the derived atomic data was given. In the present work, 
the data set has been significantly extended. First of all, we add line data 
involving the transitions from the 2$s$ shell since their wavelengths lie 
within the same UTA range. Furthermore, we determine excitation (both from 
2$s$ and 2$p$ shells) to the $n= 3, 4, 5$ shells. Finally, photoexcitation 
data for the same type of transitions in Fe\,{\sc xiv} ions are derived.

In our current work we use the configuration interaction (CI) approximation for 
atomic  calculations. Particularly, the well-established computer code 
{\sc civ3} of \citep{civ3} is exploited to determine multiconfiguration
eigenfunctions and eigenvalues of considered energy levels. After this initial
step, we derive atomic data such as electric dipole (E1) radiative transition 
wavelengths $\lambda$, oscillator strengths $f$ and transition probabilities 
(rates) $A$ for the transitions from inner $n=2$ shell of Fe\,{\sc xiv}, 
Fe\,{\sc xv} and Fe\,{\sc xvi} ions. The adopted CI method is suitable for
production of very accurate data, where the correlation effects play a vital
role in theoretical data accuracy. 

Data are generated in the relativistic Breit-Pauli approach \citep{breitp}
for all lines arising from 2$s$ and 2$p$ shell electron transitions to the 
$n =3, 4, 5$ shells. Combining these two approaches, we can reliably determine 
$LSJ$-energy levels and, consequently, the E1 transition data for 
fine-structure levels. When CI wavefunctions for the fine-structure
levels of the ground and excited states are determined, the electric dipole
transition absorption oscillator strength $f$ is given by

\begin{equation}
\label{eq:f_l}
f = 2/3\, g_i^{-1}\, \Delta E_{ij} \,S^2
\end{equation}
where $g_i = 2J_i + 1$ is the statistical weight of the initial (ground) 
level, $\Delta E_{ij}$ is the energy difference between the upper and lower
level, and $S$ is transition line strength (matrix element). The corresponding 
electric dipole line emission transition probability
$A$ (in s$^{-1}$) is given by
\begin{equation}
\label{f:A_l}
A = 2.142 \times 10^{10} \,{g_j}^{-1} \,(\Delta E_{ij})^{3} \,S^2
\end{equation}
where $g_j = 2J_j + 1$ is the statistical weight of the excited-state level $j$.

A more detailed description of the approximations used in the present work may
be found in \citet{fe1514}, where inner 2$p$ shell photoexcitation was 
considered for Fe\,{\sc xv} and Fe\,{\sc xvi}. The approach applied in the current
paper for a much wider set of lines and Fe ions is very similar. \citet{fe1514}
have considered in detail the accuracy of determined results by applying mainly
the internal indicators such as convergence of transition wavelengths with 
increasing the size of CI wavefunction expansion or agreement between length
and velocity forms of oscillator strength values. They have estimated that 
2$p - 3l$ transition wavelength accuracy was $0.2\,\%$ whereas oscillator
strength accuracy was assessed as $10\,\%$ for the lines having 
$f \geq 5 \times 10^{-5}$. The conclusions on the accuracy of the atomic 
data made by \citet{fe1514} are also applicable to the present results. 

\subsection{Fe\,{\sc xvi} lines}
\label{fe15}

The ground level of the Fe$^{15+}$ ion is $1s^22s^22p^63s$
$^2S_{1/2}^{\mathrm e}$. Its CI wavefunction expansion  
was chosen to include no more than two electrons excited from the 2$p$  and 3$s$
shells in the case of the lines arising from the 2$p$ shell, and no more 
than two electrons excited from 2$s$ and 3$s$ shells for the lines arising 
from the 2$s$ shell. The significant difference between the calculation of $n=3$
and 4 lines lies in the fact that we use correlation radial orbitals
${\bar 4}s$, ${\bar 4}p$, ${\bar 4}d$, ${\bar 4}f$ in the former case to 
improve the calculated wavelength accuracy (for more details see 
\citealp{fe1514}). In the latter case these orbitals represent real 
states while the configuration sets remain the same.

\subsubsection{Excitation from the 2p shell}
\label{fe15_2p}

Following this approach, the ground state CI wavefunction expansion for the 
transition arrays $2p - 3l$ and $2p - 4l$ was the same and included  
$2p^63l$,$2p^64l$ ($l=0,2$), 
$2p^53s3p$, $2p^53p3d$, 
$2p^53l4l^{\prime}$, $2p^54l4l^{\prime}$ ($l=0,1,2; l^{\prime}=0,1,2,3$)
even-parity configurations, coupled to the 
$^2S$, $^2P$, $^4P$, $^4D$ 
$LS$-terms. This CI expansion generates 94 $LSJ$-levels with total orbital 
quantum number $J = 1/2$. Due to selection rules for E1 transitions, the 
excited configuration with the vacancy in the 2$p$ shell can only have levels 
with $J=1/2$ and $J=3/2$. We use a CI wavefunction expansion consisting of 
odd-parity configurations
$2p^63p$, $2p^64p$, 
$2p^53l^2$ ($l=0,1,2$), $2p^53s3d$,
$2p^53l4l^{\prime}$ ($l=0,1,2;l^{\prime}=0,1,2,3$),
$2p^54l^2$($l=0,1,2,3$), $2p^54l4l^{\prime}$ ($l=0,1; l^{\prime}=2,3 $)
coupled to the $^2S$, $^2P$, $^4P$, $^4D$ $LS$-terms for $J = 1/2$ and to the 
$^4S$, $^2P$, $^4P$, $^2D$, $^4D$, $^4F$ terms for $J = 3/2$. 
Consequently, we consider 93 levels with $J = 1/2$ and 147 levels with $J = 3/2$
for the excited configurations having electrons in the outer $n = 3$ and 4 
shells and the vacancy in the inner 2$p$ shell. 

Very similar configuration sets were used to obtain atomic data for the 
transitions from the 2$p$ shell to the outer $5l$ shell. The ground state 
wavefunction expansion included
$2p^6nl$ ($n=3,4,5; l=0,2$),
$2p^53s3p$, $2p^53p3d$, 
$2p^53l5l^{\prime}$ ($l=0,1,2; l^{\prime}=0,1,2,3$),
$2p^55l5l^{\prime}$ ($l=0,1,2,3; l^{\prime}=0,1,2,3$)
even-parity configurations coupled to the 
$^2S$, $^2P$, $^4P$, $^4D$ 
$LS$-terms. This yields 95 $LSJ$-levels having $J = 1/2$. 

The same (as in $n = 3, 4$ case) terms for the excited configuration levels 
with $J = 1/2$ and $J = 3/2$ were considered, while the CI expansion included 
configurations 
$2p^6 np$ ($n=3,4,5$),  
$2p^53l^2$ ($l=0,1,2$), $2p^53s3d$,
$2p^53l5l^{\prime}$ ($l=0,1,2; l^{\prime}=0,1,2,3$)
$2p^55l^2$ ($l=0,1,2,3$), $2p^55s5d$, $2p^55p5f$. 
For  $J = 1/2$, the CI expansion included 94 levels whereas for $J = 3/2$ 
the number was 148. Levels arising from the configurations with 
5$g$ electrons were not included in the CI expansion because such  lines are 
very weak compared to other $n = 5$ transitions.

\subsubsection{Excitation from the 2s shell}
\label{fe15_2s}

A similar approach for the configuration sets to that adopted for the 2p shell 
was taken when calculating atomic data for the lines corresponding to the 
transitions from the inner 2$s$ shell to the outer $n =3, 4, 5$ shells. 
For the $n =3, 4$ lines, the ground state CI expansion included 56 levels 
with $J = 1/2$ from the configurations
$2p^6nl$ ($n=3,4; l=0,2$)
$2p^53s3p$, $2p^53p3d$,
$2s2p^63l^2$ ($l=0,1,2$),
$2s2p^63l4l^{\prime}$ ($l=0,1,2; l^{\prime}=0,1,2,3$),
$2s2p^64l^2$, ($l=0,1,2,3$),
$2s2p^64s4d$, $2s2p^64p4f$ 
coupled to the $^2S$, $^2P$, $^4P$, $^4D$ terms.
The upper-state configuration set consisted of 
$2p^63p$, $2p^64p$, $2p^64f$,
$2s2p^63s3p$, $2s2p^63p3d$,
$2s2p^6nl4l^{\prime}$ ($n=3,4; l=0,1,2; l^{\prime}=0,1,2,3$)
configurations coupled to the terms
$^2S$, $^2P$, $^4P$, $^4D$, $^6D$, $^6F$ 
for the $J = 1/2$  levels, and to the 
$^2P$, $^2D$, $^4S$, $^4P$, $^4D$, $^4F$, $^6P$, $^6D$, $^6F$, $^6G$ 
terms for $J = 3/2$. This selection produced 38 and 58 levels for $J = 1/2$ 
and 3/2, respectively.

For the lines corresponding to the transitions from the inner 2$s$ shell to the
outer $n = 5$ shell, the ground state wavefunction CI expansion included 
configurations
$2p^6nl$ ($n=3,4,5; l=0,2$),
$2p^53s3p$, $2p^53p3d$, 
$2s2p^63l^2$ ($l=0,1,2$), $2s2p^63s3d$,
$2s2p^63l5l^{\prime}$ ($l=0,1,2; l^{\prime}=0,1,2,3$), 
$2s2p^65l^2$ ($l=0,1,2$), $2s2p^65s5d$, $2s2p^65p5f$ 
coupled to the terms $^2S$, $^2P$, $^4P$, $^4D$.
This generates a total of 57 levels with $J = 1/2$. 
For the upper state configuration set we included configurations 
$2p^6nl$ ($n=3,4,5; l=1,3$),
$2s2p^63s3p$, $2s2p^63p3d$,
$2s2p^6nl5l^{\prime}$, ($n=3,4; l=0,1,2; l^{\prime}=0,1,2,3$)
which were coupled to the terms 
$^2S$, $^2P$, $^4P$, $^4D$, $^6D$, $^6F$, 
producing 39 levels with $J = 1/2$. For $J = 3/2$ we considered the same 
terms as in the $n = 4$ lines case, producing 59 levels.

\subsubsection{Transition data}
\label{td15}

We employ multiconfiguration wavefunctions determined after the Hamiltonian
matrix diagonalization to calculate transition wavelengths $\lambda$, 
oscillator strengths $f$, and transition rates $A$ for excitation from the 
inner $n=2$ shell to the outer $n=3, 4, 5$ shells. 
Calculated line data cover the wavelength region 
$\lambda= 12.606 - 14.096$\,\AA \,\, for $2s - 3l$ transitions (a total of 17 
lines), $\lambda= 10.145 - 11.221$\,\AA \,\, for $2s - 4l$ transitions (46 
lines) and $\lambda= 9.545 - 10.333$\,\AA \,\, for $2s - 5l$ transitions (46 
lines). Considering transitions from the 2$p$ shell, the wavelength range was 
$\lambda= 13.594 - 17.246$\,\AA \,\, for the excitation of the $n=3$ shell 
(45 lines), $\lambda= 11.316 - 12.934$\,\AA \,\, for the $n=4$ shell (111
lines), and $\lambda= 10.525 - 11.689$\,\AA \,\, for the $n=5$ shell (111 
lines).

%%%%%%%%%%%%%%%%%% T1 %%%%%%%%%%%%%%%%%%%%%%%%%%%%%%
\begin{deluxetable}{ccccrr}
\tabletypesize{\scriptsize}
\tablecolumns{6} 
\tablewidth{0pt}
%\tablenum{1}
\tablecaption{
Transitions $2p - n_jd$ and $2s - n_jp $ from the $L$ shell to $n_j = 3,4,5$ 
shells in Fe$^{15+}$. Column $n_il_i$ denotes the initial shell, $n_j$ the 
final shell for optical electron transition, $g_i$ and $g_j$ the statistical 
weights for the initial and final levels, $\lambda$ the line wavelength 
(in \AA) and $f_{ij}$ the absorption oscillator strength. 
\label{tab_fe15short}
} 
\tablehead{ 
\colhead{$n_il_i$} &
\colhead{$n_j$} & 
\colhead{g$_i$} &
\colhead{g$_j$} & 
\colhead{$\lambda $(\AA)} &
\colhead{$f_{ij}$}
}
\startdata 
2p	&	3	&	2	&	4	&	15.153	&	0.841	\\
2p	&	3	&	2	&	2	&	15.083	&	0.795	\\
2p	&	3	&	2	&	4	&	14.996	&	0.697	\\
2s	&	3	&	2	&	4	&	13.951	&	0.291	\\
2p	&	4	&	2	&	4	&	12.336	&	0.230	\\
2p	&	4	&	2	&	4	&	12.469	&	0.227	\\
2p	&	3	&	2	&	4	&	15.408	&	0.158	\\
2p	&	4	&	2	&	2	&	12.330	&	0.140	\\
2s	&	3	&	2	&	2	&	13.971	&	0.131	\\
2p	&	3	&	2	&	4	&	15.321	&	0.119	\\
2p	&	3	&	2	&	2	&	15.384	&	0.112	\\
2p	&	5	&	2	&	4	&	11.495	&	0.105	\\
2p	&	4	&	2	&	2	&	12.485	&	0.104	\\
\enddata 
\tablecomments {Table~\ref{tab_fe15short} is published in its entirety in the 
electronic edition of the {\it Astrophysical Journal}.  Only a portion is 
shown here for guidance regarding its form and content.}
\end{deluxetable}
%%%%%%%%%%%%%%%%%%%%%%%%%%%%%%%%%%%%%%%%%%%%%%%%%%%%

In Table~\ref{tab_fe15short} we present data for Fe\,{\sc xvi} 
transitions from the 2$s$ and 2$p$ shells to $n=3,4,5$. We include
only transitions with $f \geq 0.1$, whereas the more complete 
online version of the table contains lines having $f \geq 0.0001$. Weaker lines
are not presented since they do not affect the ionization balance nor
the final spectra. More details on data trimming are given in 
Sect.~\ref{trim}.

%%%%%%%%%%%%%F1 %%%%%%%%%%%%%%%%%%%%%%%%%%%%%
\begin{figure}[!t]
%\plottwo{fe15.bw.eps}{fe15.eps}
\epsscale{.50}
%\plotone{fe15.bw.eps}
%\plotone{fe15.eps}
\includegraphics[scale=0.65]{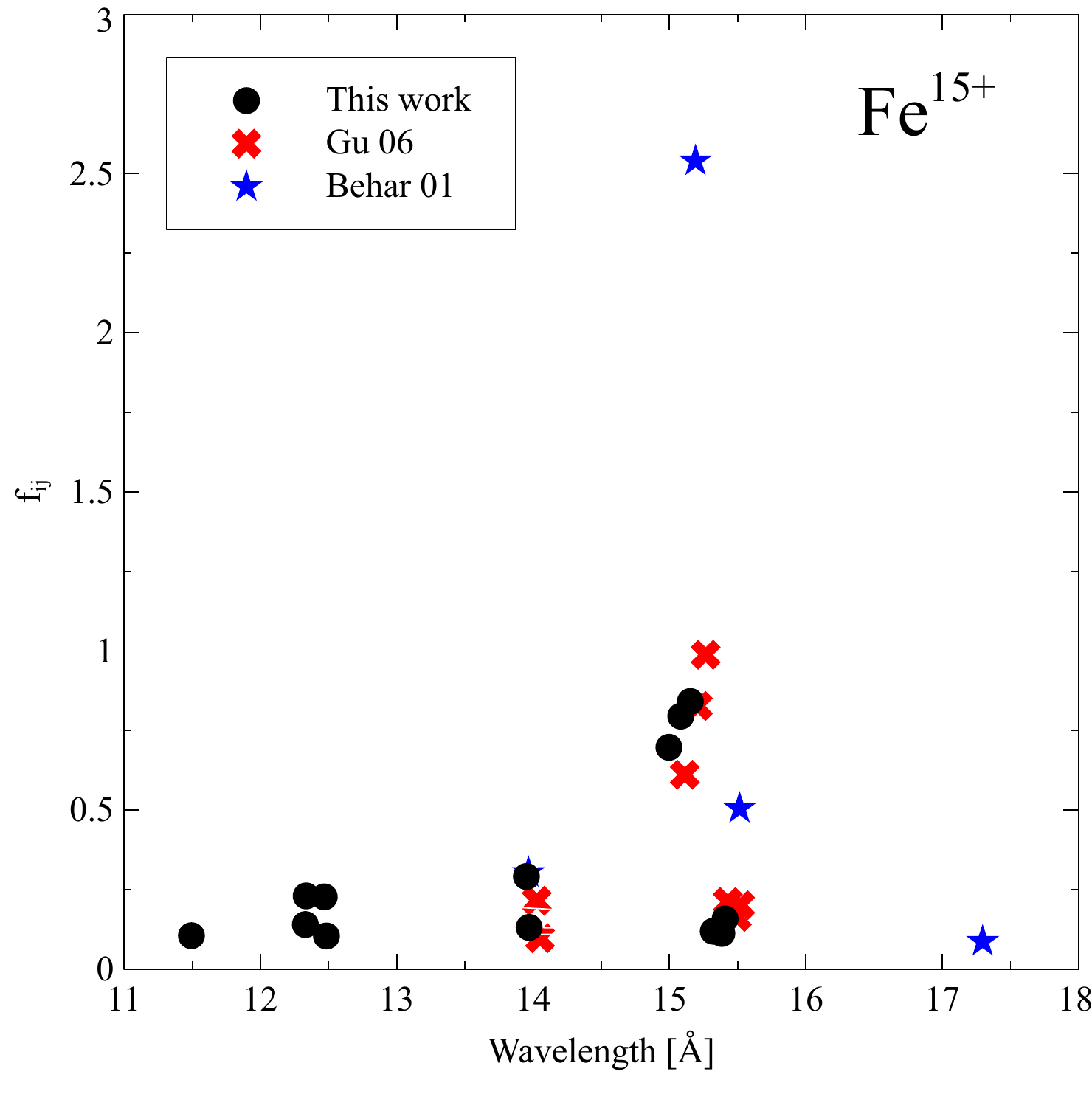}
\caption{A comparison of the wavelengths $\lambda$ and the absorption 
oscillator strengths $f_{ij}$ for the strongest lines in Fe\,{\sc xvi}.
Dots represent our data, crosses are the results of \citet{Gu2006} and 
stars are those from \citet{Behar2001}. 
The data at $\lambda=15$\AA\, represent $2p - 3d$ UTA, 
at $\lambda=14$\AA\, represent $2s - 3p$ UTA, at $\lambda=12.5$\AA\,
represent $2p - 4d$ UTA and at $\lambda=11.5$\AA\, represent $2p - 5d$ UTA.
The \citet{Behar2001} data are given for an $f$-value averaged wavelength 
$\lambda_{\mathrm{av}}$ and the total oscillator strength $\sum f_{ij}$ 
of the line group.
See the electronic edition of the Journal for a color version 
of this figure.\label{fig15}}
\end{figure}
%%%%%%%%%%%%%%%%%%%%%%%%%%%%%%%%%%%%%%%%%%%

In Fig.~\ref{fig15} we compare the strongest lines from our calculation with
data derived using a many-body perturbation theory (MBPT, \citealt{Gu2006}), 
and with the \citet{Behar2001} results obtained using the multiconfiguration 
relativistic HULLAC computer package. 
The \citet{Behar2001} data are given for an $f$-value averaged wavelength 
$\lambda_{\mathrm{av}}$ and the total oscillator strength $\sum f_{ij}$ of the
line group, and therefore there is a sizable difference in the number of lines 
and in the values of $f$. Note that the point at $\lambda=17.296$\AA\, 
represents transitions from the 2$p$ to 3$s$ shells which are not considered in
our calculations. When comparing our results with those of \citet{Gu2006}, one 
can see quite  good agreement both for wavelengths and oscillator strengths. 
There is some systematic shift towards shorter wavelengths in our data, as 
discussed in \citet{fe1514}. In addition, a significant difference with Gu et 
al. (2006) is the appearance in our results of a group of lines at 
$\lambda = 11.5 - 12.5\,$\AA\, which represent transitions to $n=4$ and $n=5$ 
states. Some of these have large $f$-values and are important to include in 
atomic models.

The primary goal of the calculations presented here is to create an extensive 
set of inner-shell transition data to ensure that the total excitation and 
ionization rate is computed as accurately as possible.  In a later section we 
show how this new data changes the ionization distribution of gas near an AGN.
The line wavelengths we quote, which come from our atomic structure calculation,
are no better or worse than the wavelengths quoted in previous studies.  
Indeed, the scatter in the wavelengths represents the uncertainty.  Extensive 
sets of laboratory measurements of energy levels would be needed to improve 
the level energies and resulting wavelengths.  Although a few experiments have
been done (\cite{brown2001}  and \cite{simon2010}), the extensive laboratory 
data needed to significantly improve the line wavelengths do not now exist. 

\subsection{Fe\,{\sc xv} lines}
\label{fe14}

The ground level of the Fe$^{14+}$ ion is $1s^22s^22p^63s^2$ 
$^1S_{0}^{\mathrm e}$. Its CI wavefunction expansion 
was chosen to include no more than two electrons virtually excited from the 
2$p$ and 3$s$ shells in the case of the lines arising from the 2$p$ shell, 
and no more than two electrons excited from the 2$s$ and 3$s$ shells for the 
lines arising from the 2$s$ shell. We adopt the same configuration set both 
for the $2p-3l$ and the $2p-4l$ line calculations, but different radial orbital 
sets are used.
For the $n=3$ line calculation, we employ the correlation radial orbitals 
${\bar 4}s$, ${\bar 4}p$, ${\bar 4}d$, ${\bar 4}f$, whereas the corresponding 
radial orbitals represent real states in the $n=4$ line calculations.  
This method contributes to an increased accuracy of the calculated wavelengths 
for $n=3$ lines which are much stronger than the $n=4$ ones.

\subsubsection{Excitation from the 2p shell}
\label{fe14_2p}

For excitation from 2$p$ to the $n=3,4$ shells, a CI wavefunction expansion
was made of the
$2p^63l^2$ ($l=0,1,2$), $2p^63s3d$,
$2p^53s^23p$, $2p^53p^3$, $2p^53p3d^2$, $2p^53s3p3d$
configurations containing two or three electrons within the $n = 3$ shell.
The even-parity configurations
$2p^63l4l^{\prime}$ ($l=0,1,2; l^{\prime}= 0,1,2,3$),
$2p^53l^24l^{\prime}$,($l=0,1,2; l^{\prime}= 1,3$),
$2p^53l3l^{\prime}4l^{\prime\prime}$
($l=0,1; l^{\prime}= 0,1,2; l^{\prime\prime} = 0,1,2,3$)
with a single electron virtually excited to the $n = 4$ shell
were added to that set as well as even-parity configurations
$2p^64l^2$ ($l=0,1,2,3$), 
$2p64s4d$, $2p^64p4f$,
$2p^53p4l^2$ ($l=0,1,2,3$),
$2p^53l4l^{\prime}4l^{\prime\prime}$ 
($l=0,1,2; l^{\prime}= 0,1,2; l^{\prime\prime} = 1,2,3$)
with two electrons excited to the $n = 4$ shell. 
In the case of inner 2$p$ shell excitation, we consider the 
$^1S^{\mathrm e}$, $^3P^{\mathrm e}$ and $^5D^{\mathrm e}$ terms which 
produce the $J = 0$ fine-structure levels. Construction of such a basis 
produces 408 configuration state functions (CSFs) in the CI wavefunction 
expansion for the lower state of the Fe$^{14+}$ ion.

Due to  the electric dipole transition selection rules, we have to consider 
fine-structure levels with $J = 1$ in the upper (odd) state of this ion. 
In the same way as for the lower (even) state, we construct three subsets of 
configurations in wavefunction expansion for the upper (odd) state. The CI 
wavefunction expansion consists of the 
$2p^63s3p$, $2p^63p3d$, 
$2p^53s^23d$, $2p^53s3p^2$, $2p^53s3d^2$, $2p^53p^23d$, $2p^53d^3$
odd-parity configurations which have two or three electrons in the $n = 3$ 
shells. The odd-parity configurations
$2p^63l4l^{\prime}$ ($l=0,1,2; l^{\prime}=0,1,2,3$),
$2p^53l^24l^{\prime}$ ($l=0,1,2; l^{\prime}=0,2$),
$2p^53l3l^{\prime}4l^{\prime\prime}$
($l=0,1; l^{\prime}= 0,1,2; l^{\prime\prime} = 0,1,2,3$)
with all possible electron distributions having a single electron
virtually excited to the $n = 4$ shell and the same parity configurations
with two electrons in open $n = 4$ shells
$2p^64l4l^{\prime}$ ($l=0,1,2; l^{\prime}=1,2,3$),
$2p^53l4l^{\prime\,2}$ ($l=0,2; l^{\prime}=0,1,2,3$), 
$2p^53l4l^{\prime}4l^{\prime\prime}$ 
($l=0,1,2; l^{\prime}=0,1,2; l^{\prime\prime}=1,2,3$
were added to that expansion.

In this construction of the CI wavefunction basis, we restricted our set by 
allowing no more than two electrons in the excited $n = 4$ shell. Here we 
consider the $^3S^{\mathrm o}$, $^1P^{\mathrm o}$, $^3P^{\mathrm o}$, 
$^5P^{\mathrm o}$, $^3D^{\mathrm o}$, $^5D^{\mathrm o}$, $^5F^{\mathrm o}$
terms which can produce fine-structure levels with $J = 1$. As a result,
we have the CI wavefunction expansion with 1094 CSFs for the upper states of 
the Fe$^{14+}$ ion.

A very similar configuration set was applied when considering $2p - 5l$ 
transitions. In this instance, the configurations with $n=4$ electrons were
replaced by configurations with $n=5$. The only significant difference
was the addition of the 
$2p^63l4l^{\prime}$ ($l=0,1,2; l^{\prime}=0,1,2,3$)
even-parity configurations for in CI expansion of the lower state of Fe$^{14+}$,
and the $2p^63l4l^{\prime}$ ($l=0,1,2; l^{\prime}=0,1,2,3$)
odd-parity configurations for the excited state. Configurations with 5$g$ 
electrons were not included in the CI wavefunction expansion because 
transitions to such configurations are very weak. The total number of 
fine-structure levels with $J=0$ in this case was 413 for the lower 
configuration and 1107 for the upper configuration levels with
$J=1$. 

\subsubsection{Excitation from the 2s shell}
\label{fe14_2s}

While determining the transition array $2s - 3l, 4l$ lines, for the CI expansion
we have adopted the same subset (as in $2p - 3l, 4l$ case) of configurations 
with closed $2s^2$ and $2p^6$ electron shells and complemented it with 
configurations having one electron virtually excited from the 2$s$ shell. 
Specifically for the ground level, the even-parity configurations
$2s2p^63s^23d$, $2s2p^63p^23d$, $2s2p^63s3p^2$, 
$2s2p^63s3d^2$, $2s2p^63d^3$,
$2s2p^63l^24l^{\prime}$ ($l=0,1,2; l^{\prime}=0,2$), 
$2s2p^63l3l^{\prime}4l^{\prime\prime}$
($l=0,1; l^{\prime}=1,2; l^{\prime\prime}=0,1,2,3$),
$2s2p^63l4l^{\prime\,2}$ ($l=0,2; l^{\prime}=0,1,2,3$), 
$2s2p^63l4l^{\prime}4l^{\prime\prime}$ 
($l,l^{\prime}=0,1,2; l^{\prime\prime}=1,2,3$), 
were included in our CI wavefunction expansion. These configurations were
coupled in the $^1S^{\mathrm e}$, $^3P^{\mathrm e}$ and $^5D^{\mathrm e}$ 
non-relativistic $LS$-terms and produced 165 fine-structure levels with
total orbital quantum number $J=0$.

The construction of the configuration set for the upper state is based on 
the same principles. We employ configurations with closed $2s^2$ and $2p^6$ 
electron shells from the $2p - 3l, 4l$ array calculation and add the 
configurations with an open 2$s$ shell. Following this method, we include 
the configurations 
$2s2p^63s^23p$, $2s2p^63p3d^2$, $2s2p^63p^3$, $2s2p^63s3p3d$ 
with the 2$s$ electron virtually excited to the $n=3$ shell, 
the odd-parity configurations
$2s2p^63l^24l^{\prime}$ ($l=0,1,2; l^{\prime}=1,3$), 
$2s2p^63l3l^{\prime}4l^{\prime\prime}$
($l=0,1,2; l^{\prime}=1,2; l^{\prime\prime}=0,1,2,3$)
with the 2$s$ electron virtually excited to the $n=4$ shell, 
and the odd-parity configurations with two electrons in the $n=4$ shell
$2s2p^63l4l^{\prime}4l^{\prime\prime}$, 
($l=0,1,2; l^{\prime}=0,1,2; l^{\prime\prime}=1,2,3$).
Coupled to the $^3S^{\mathrm o}$, $^1P^{\mathrm o}$, $^3P^{\mathrm o}$, 
$^5P^{\mathrm o}$, $^3D^{\mathrm o}$, $^5D^{\mathrm o}$, 
$^5F^{\mathrm o}$ terms, these configurations generate total 417 
fine-structure levels with $J=1$.

For the transition array $2s - 5l$, the wavefunction CI expansion was similar 
to that for the $2s - 4l$ lines with the $n=4$ shell electrons replaced by 
$n=5$. The only addition was even-parity configurations
$2p^63l4l^{\prime}$ ($l=0,1,2; l^{\prime}=0,1,2,3$
with one electron in the $4l$ shell, included in the lower-state CI 
wavefunction expansion set, and the odd-parity configurations
$2p^63l4l^{\prime}$ ($l=0,1,2; l^{\prime}=0,1,2,3$
in the upper-state wavefunction expansion. This CI expansion selection
produces 170 fine-structure levels with $J=0$ for the ground state and 430
levels with $J=1$ for the excited state.

\subsubsection{Transition data}
\label{td14}

Similarly to the Fe\,{\sc xvi} case, we obtained multiconfiguration 
wavefunctions to produce photoexcitation data for the Fe\,{\sc xv} UTA lines.
The calculated line data cover the wavelength region 
$\lambda= 11.592 - 14.199$\,\AA \,\, for 2$s$ - $3l$ transitions 
(a total of 35 lines), 
$\lambda= 9.620 - 11.423$\,\AA \,\, for 2$s$ - $4l$ transitions (147 lines) 
and $\lambda= 9.545 - 10.333$\,\AA \,\, for 2$s$ - $5l$ transitions (147 
lines). Considering transitions from the 2$p$ shell, the wavelength range was 
$\lambda= 12.475 - 16.266$\,\AA \,\, for the excitation of the $n=3$ shell 
(88 lines), $\lambda= 10.630 - 13.244$\,\AA \,\, for the $n=4$ shell (384
lines), and $\lambda= 9.998 - 12.017$\,\AA \,\, for the $n=5$ shell (384 lines).

In Table~\ref{tab_fe14short} we present a sample of the line data for 
Fe\,{\sc xv}. Transition wavelengths $\lambda$ and oscillator strengths f$_{ij}$ 
for excitation from the 2$s$ and 2$p$ shells to the outer $n= 3, 4, 5$ shells 
are presented. Only a small portion of lines with $f \geq 0.4$ is included, 
with the more complete online version of the table containing all lines which 
have $f \geq 0.0001$. It is evident from Table~\ref{tab_fe14short} that the 
lines representing transitions to the $n=3,4$ shells are sufficiently strong as
to affect spectral formation, and therefore they cannot be excluded from UTA 
line data sets.

%%%%%%%%%%%%%%% T2 %%%%%%%%%%%%%%%%%%%%%%%%%%%%%%%%%
\begin{deluxetable}{ccccrr}
\tabletypesize{\scriptsize}
\tablecolumns{6} 
\tablewidth{0pt}
%\tablenum{1}
\tablecaption{
Transitions $2p - n_jd$ and $2s - n_jp $ from the $L$ shell to the $n_j = 3, 4, 5$ 
shells in Fe$^{14+}$. Column $n_il_i$ denotes the initial shell, $n_j$ the final
shell for optical electron transition, $g_i$ and $g_j$ the statistical weights 
for initial and final levels, $\lambda$ the line wavelength (in \AA) and 
$f_{ij}$ the absorption oscillator strength.
\label{tab_fe14short}
} 
\tablehead{ 
\colhead{$n_il_i$} &
\colhead{$n_j$} & 
\colhead{g$_i$} &
\colhead{g$_j$} & 
\colhead{$\lambda $(\AA)} &
\colhead{$f_{ij}$}
}
\startdata 
2p&	3&	1&	3&	15.227&	1.790\\   
2p&	3&	1&	3&	15.479&	0.600\\
2s&	3&	1&	3&	14.121&	0.405\\
2p&	4&	1&	3&	12.759&	0.359\\
2p&	4&	1&	3&	12.619&	0.296\\
2p&	5&	1&	3&	11.820&	0.177\\
2p&	3&	1&	3&	15.374&	0.138\\
2p&	5&	1&	3&	11.696&	0.117\\
2s&	4&	1&	3&	11.407&	0.097\\
2p&	4&	1&	3&	12.615&	0.065\\
2p&	3&	1&	3&	16.008&	0.060\\
2p&	3&	1&	3&	15.779&	0.041\\
2s&	3&	1&	3&	14.190&	0.040\\
\enddata 
\tablecomments {Table \ref{tab_fe14short} is published in its entirety in the 
electronic edition of the {\it Astrophysical Journal}.  Only a portion is 
shown here for guidance regarding its form and content.}
\end{deluxetable} 
%%%%%%%%%%%%%%%%%%%%%%%%%%%%%%%%%%%%%%%%%%%%%%%

%%%%%%%%%%%% F2 %%%%%%%%%%%%%%%
\begin{figure}[!t]
%\plotone{fe14.bw.eps}
%\plotone{fe14.eps}
\includegraphics[scale=0.65]{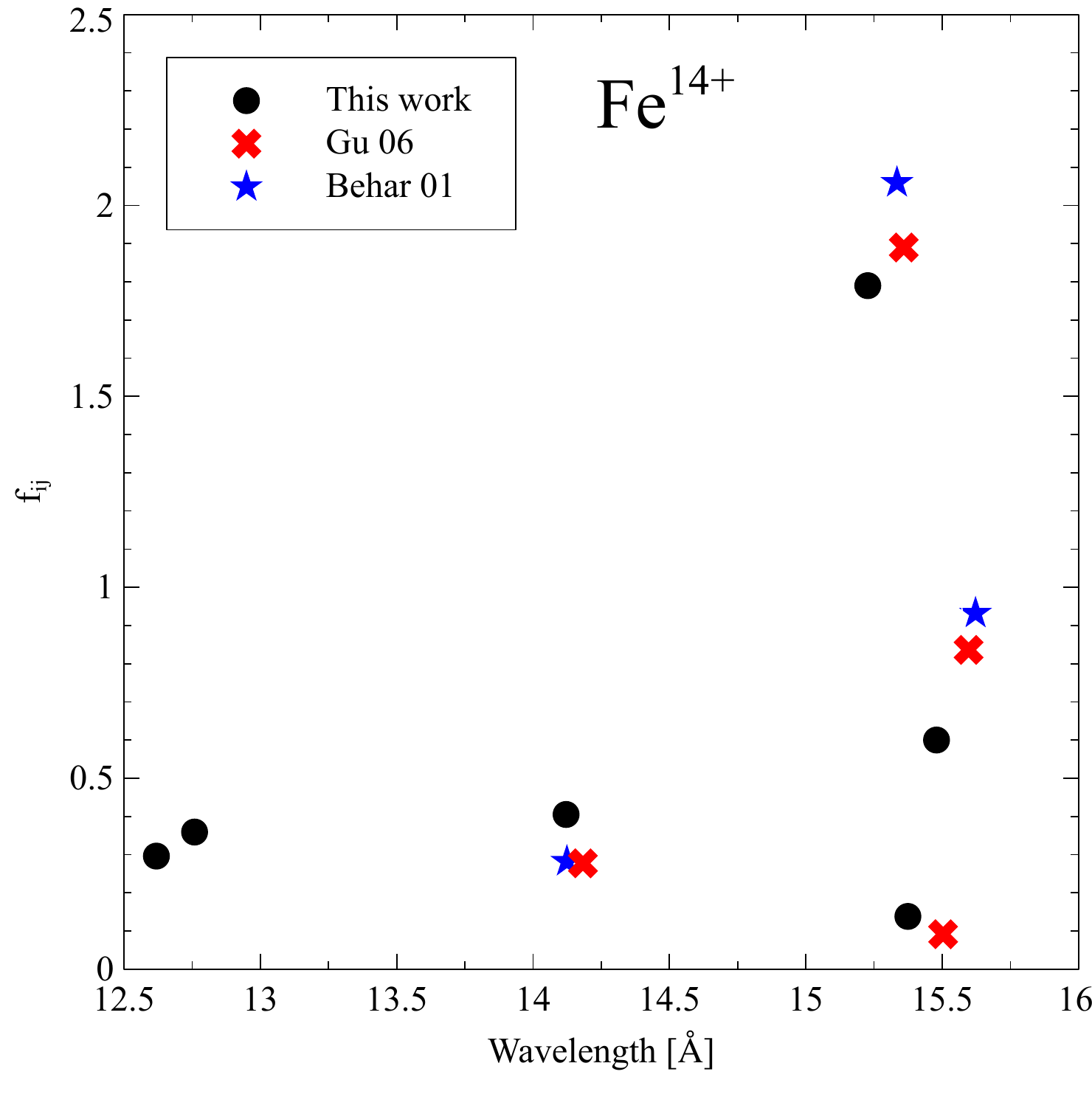}
\caption{A comparison of the wavelengths $\lambda$ and absorption oscillator
strengths $f_{ij}$ for the strongest lines in Fe\,{\sc xv}.
Dots represent our data, crosses the results of
\citet{Gu2006} and stars those from \citet{Behar2001}. 
The data at $\lambda=15.5$\AA\, represent $2p - 3d$ UTA, 
those at $\lambda=14.1$\AA\, represent $2s - 3p$ UTA, and those at $\lambda=12.7$\AA\,
represent $2p - 4d$ UTA.
The \citet{Behar2001} data are given for an $f$-value averaged wavelength 
$\lambda_{\mathrm{av}}$ and the total oscillator strength $\sum f_{ij}$ 
of the line group.
See the electronic edition of the Journal for a color version 
of this figure.\label{fig14}}
\end{figure}
%%%%%%%%%%%%%%%%%%%%%%%%%%%

Fig. \ref{fig14} compares the strongest lines from our calculations with
the results derived using the MBPT approach \citep{Gu2006} and with the 
relativistic data from \citet{Behar2001}. In general, the agreement between all
three sets of data is satisfactory. There is a systematic shift towards shorter 
wavelengths in our data set as noted for the Fe\,{\sc xvi} lines. 

\subsection{Fe\,{\sc xiv} lines}
\label{fe13}

The ground configuration of the Fe$^{13+}$ ion is $1s^22s^22p^63s^23p$. 
It has two fine-structure levels, $^2P_{1/2}^{\mathrm o}$ and 
$^2P_{3/2}^{\mathrm o}$, separated by 2.33741 eV. Consequently, while
considering electric-dipole transitions, we must determine energy levels
with total angular momenta $J=1/2, 3/2, 5/2$ of the excited configuration of
the Fe$^{13+}$ ion.

We apply real radial orbitals for $1s$, $2s$, $2p$, $3s$, $3p$, $3d$ electrons 
and correlation ones for the virtually excited $4s$, $4p$, $4d$, $4f$ electrons.
This is consistent with the methods used to determine transition data for 
Fe\,{\sc xvi} and Fe\,{\sc xv} (see \citet{fe1514}), for photoexcitation from 
the $2p$ and $2s$ shells of the ground configuration to the levels of the 
excited $n=3$ configurations.
Furthermore, for the excitation of the $n=4$ and $n=5$ lines, we adopt only 
real radial orbitals to describe $4l$ and $5l$ ($l=0,1,2,3$) electrons. 

\subsubsection{Excitation from the 2p shell}
\label{fe13_2p}

We employ the same configuration set both for calculations of $2p - 3l$ and 
$2p - 4l$ transition data. The CI expansion for the lower state includes 
configurations which have one or two electrons virtually excited from the $n=3$ 
shell. These are complemented with configurations which have the 
2$p$ electron moved to the $n=3$ shell. The CI wavefunction expansion 
for the ground state was composed of the configurations with three or four 
electrons in $n=3$ shell
$2p^63s^23p$, $2p^63p3d^2$, $2p^63p^3$, $2p^63s3p3d$, 
$2p^53s^23p^2$, $2p^53p^23d^2$, $2p^53s3d^3$, $2p^53p^4$, $2p^53d^4$, 
$2p^53s3p^23d$, 
the odd-parity configurations with one electron in virtually-excited $n=4$ shell
$2p^63l^24l^{\prime}$ ($l=0,1,2; l^{\prime}=1,3$)
$2p^63l3l^{\prime}4l^{\prime\prime}$ 
$l=0,1; l^{\prime}=1,2; l^{\prime\prime}=0,1,2,3$),
and the odd-parity configurations with two electrons in $n=4$ shell
$2p^63l4l^{\prime}4l^{\prime\prime}$
$l=0,1,2; l^{\prime}=0,1,2; l^{\prime\prime}=1,2,3$),
$2p^63p4l^2$ ($l=0,1,2,3$).
In the case of the $J=1/2$ levels, we consider the 
$^2S$, $^2P$, $^4P$, $^4D$, $^6D$, $^6F$ 
odd-parity $LS$-terms making up 336 CSFs in the CI wavefunction expansion, 
while for the $J=3/2$ fine-structure levels, the odd-parity $LS$-terms
$^2P$, $^2D$, $^4S$, $^4P$, $^4D$, $^4F$, $^6P$, $^6D$, $^6F$, $^6G$ 
produce 553 CSFs. 

For the upper state levels, the CI wavefunction expansion for the 
$2p - 4l$ lines calculation is constructed from even-parity configuration 
sets. 
The configurations  with electrons in $n=3$ shell
$2p^63s^23d$, $2p^63s3p^2$, $2p^63s3d^2$, $2p^63p^23d$,  $2p^63d^3$
the even-parity configurations with one electron in $n=4$ shell
$2p^63l^24l^{\prime}$ ($l=0,1,2; l^{\prime}=0,2$),
$2p^63l3l^{\prime}4l^{\prime\prime}$, 
($l=0,1; l^{\prime}=1,2; l^{\prime\prime}=0,1,2,3$)
the even-parity configurations with two electrons virtually excited
into the $n=4$ shell, namely
$2p^63l4l^{\prime\,2}$ ($l=0,2; l^{\prime}=0,1,2,3$),
$2p^63l4l^{\prime}4l^{\prime\prime}$ 
($l=0,1,2; l^{\prime}=0,1,2; l^{\prime\prime}=1,2,3$)
and the even-party configurations with a vacancy in the inner 2$p$ shell:
$2p^53s^23p3d$, $2p^53s3p3d^2$, $2p^53s3p^3$, $2p^53p3d^3$, $2p^53p^33d$, 
$2p^53s^23l4l^{\prime}$ ($l=1,2; l^{\prime}=0,1,2,3$),
$2p^53s3l^{\prime\,2}4l^{\prime\prime}$ 
($l=0,1; l^{\prime}=1,2; l^{\prime\prime}=0,1,2,3$),
$2p^53l^34l^{\prime}$ ($l=1,2; l^{\prime}=0,1,2,3$),
$2p^53l^24l^{\prime}4l^{\prime\prime}$, 
($l=0,1,2; l^{\prime}=0,1,2; l^{\prime\prime}=1,2,3$),
$2p^53l3l^{\prime}4l^{\prime\prime\, 2}$
($l=0,1; l^{\prime}=1,2; l^{\prime\prime}=0,1,2,3$),
$2p^53l3l^{\prime}4l^{\prime\prime}4l^{\prime\prime\prime}$
($l=0,1; l^{\prime}=1,2; l^{\prime\prime}=0,1,2; l^{\prime\prime\prime}=1,2,3$).

These configurations are bound to the 
$^2S$, $^2P$, $^4P$, $^4D$, $^6P$, $^6D$, $^6F$ 
$LS$-terms for $J=1/2$ and produce 4099 CSFs in the CI wavefunction expansion. 
The terms
$^2P$, $^2D$, $^4S$, $^4P$, $^4D$, $^4F$, $^6P$, $^6D$, $^6F$, $^6G$
generate a total of 7049 fine-structure levels for $J = 3/2$. 
For the total angular momentum $J=5/2$, the above configurations are bound to the
$^2D$, $^2F$, $^4P$, $^4D$, $^4F$, $^4G$, $^6S$, $^6P$, $^6D$, $^6F$, $^6G$, $^6H$ 
non-relativistic terms and produce 8211 levels.

The CI expansion for the lower state in the case of $2p - 5l$ consists of 
configuration sets similar to those used in $2p - 3l, 4l$ lines calculation.
However, the four configuration complexes with $4l$ ($l= 0,1,2,3$) electrons 
are replaced by the configurations with $5l$ electrons. Furthermore, the CI
wavefunction expansion is extended by an additional set of the odd-parity
configurations
$2p^63l^24l^{\prime}$ ($l=0,1,2; l^{\prime}=1,3$) and 
$2p^63l3l^{\prime}4l^{\prime\prime}$
($l=0,1; l^{\prime}=1,2; l^{\prime\prime}=0,1,2,3$)
with one electron in an outer $n=4$ shell. The same (as in case of transitions
to $n=3, 4$) non-relativistic $LS$-terms produce 359 CSFs in the CI wavefunction
expansion for $J=1/2$, and 591 CSFs for $J=3/2$. 

Similarly, the configuration sets with $4l$ electrons are replaced by the 
configuration sets with $5l$ electrons for the upper states. 
This set is extended with the even-parity configurations having one electron in 
the $n=4$ shell:
$2p^63l^24l^{\prime}$ ($l=0,1,2; l^{\prime}=0,2$)
and
$2p^63l3l^{\prime}4l^{\prime\prime}$ 
($l=0,1; l^{\prime}=1,2; l^{\prime\prime}=0,1,2,3$).
As in the previous case,  even-parity $LS$-terms produce 4326, 7421 and 8612 CSFs 
in the CI wavefunction expansion for  total angular momentum $J=1/2$, 3/2 and 5/2, 
respectively.

\subsubsection{Excitation from the 2s shell}
\label{fe13_2s}

For the lines representing transitions from the inner $2s$ shell to the valence 
$n=3$ or $n=4$ shells, the CI wavefunction expansion configurations for the 
lower state consists of the same configuration complexes for both cases. 
The first are configurations with outer electrons in $n=3$ shell
$2p^63s^23p$, $32p^6p^3$, $2p^63p3d^2$, $2p^63s3p3d$ and those with $2s$ vacancy
$2s2p^63s^23p3d$, $2s2p^63s3p3d^2$, $2s2p^63s3p^3$, 
$2s2p^63p^33d$, $2s2p^63p3d^3$. 
These configurations are complemented with odd-parity configurations with one 
and two electrons in the $4l$ shell:
$2p^63l^24l^{\prime}$ ($l=0,1,2; l^{\prime}=1,3$),
$2p^63l3l^{\prime}4l^{\prime\prime}$
($l=0,1; l^{\prime}=1,2; l^{\prime\prime}=0,1,2,3$),
$2p^63l4l^{\prime}4l^{\prime\prime}$
($l=0,1,2; l^{\prime}=0,1,2; l^{\prime\prime}=1,2,3$),
$2p^63p4l^2$ ($l=0,1,2,3$),
$2s2p^63l^23l^{\prime}4l^{\prime\prime}$ 
($l=0,1; l^{\prime}=1,2; l^{\prime\prime}=0,1,2,3$),
$2s2p^63l3l^{\prime\, 2}4l^{\prime\prime}$
($l=0,1; l^{\prime}=1,2; l^{\prime\prime}=0,1,2,3$),
$2s2p^63l^34l^{\prime}$ ($l=1,2; l^{\prime}=0,1,2,3$),
and
$2s2p^63s3p3d4l$ ($l=0,2$).

In the case of the $J=1/2$ levels, the above configurations are bound to the  
$^2S$, $^2P$, $^4P$, $^4D$, $^6D$, $^6F$ odd-parity $LS$-terms which yield a 
total of 660 CSFs in the CI wavefunction expansion. For the $J=3/2$ levels 
represented by the
$^2P$, $^2D$, $^4S$, $^4P$, $^4D$, $^4F$, $^6P$, $^6D$, $^6F$, $^6G$ 
odd-parity $LS$-terms, the wavefunction expansion consists of 1102 CSFs.

The wavefunction expansion for the lower state in the case of the $2s - 5l$ 
transition array is constructed in a similar way by replacing all the 
configurations containing $4l$ ($l=0,1,2,3$) electrons with configurations 
containing $5l$ electrons. Furthermore, a complex of configurations with
one electron in the $4l$ shell is added. This extends CI expansion by including
odd-parity configurations
$2p^63l^24l^{\prime}$ ($l=0,1,2; l^{\prime}=1,3$) and
$2p^63l3l^{\prime}4l^{\prime\prime}$
($l=0,1; l^{\prime}=1,2; l^{\prime\prime}=0,1,2,3$).
The same (as for the $n=4$ lines) $LS$-terms are considered, and the CI 
wavefunction expansion gives rise to 716 and 1193 CSFs for the $J=1/2$ and 
$J=3/2$ levels, respectively.

While considering the upper state of Fe$^{13+}$, the CI wavefunction expansion 
for  $2s - 3l, 4l$ lines is constructed of several configuration complexes, 
representing configurations with one or two electrons virtually excited
from the ground configuration. We have included configurations
$2p^63s^23d$, $2p^63s3p^2$, $2p^63s3d^2$, $2p^63p^23d$, $2p^63d^3$
with three outer electrons in $n=3$ shell, the even-parity configurations
$2p^63l^24l^{\prime}$ ($l=0,1,2; l^{\prime}=0,2$),
$2p^63l3l^{\prime}4l^{\prime\prime}$
($l=0,1; l^{\prime}=1,2; l^{\prime\prime}=0,1,2,3$),
$2p^63l4l^{\prime\, 2}$ ($l=0,2; l^{\prime}=0,1,2,3$)
with one electron in outer $n=4$ shell,
$2p^63l4l^{\prime}4l^{\prime\prime}$ 
($l=0,1,2; l^{\prime}=0,1,2; l^{\prime\prime}=1,2,3$)
with two electrons in $n=4$ shell. On top of that, we have extended the CI
expansion by adding the even-parity configurations with a vacancy in the $2s$ 
shell:
$2s2p^63l^23l^{\prime\, 2}$ ($l=0,1; l^{\prime}=1,2$), 
$2s2p^63s3p^23d$, $2s2p^63s3d^3$, $2s2p^63p^4$, $2s2p^63d^4$,  
$2s2p^63l^23l^{\prime}4l^{\prime\prime}$
($l=0,1; l^{\prime}=1,2; l^{\prime\prime}=0,1,2,3$),
$2s2p^63l3l^{\prime\, 2}4l^{\prime\prime}$
($l=0,1; l^{\prime}=1,2; l^{\prime\prime}=0,1,2,3$),
$2s2p^63l^34l^{\prime}$ ($l=1,2; l^{\prime}=0,1,2,3$),
$2s2p^62s2p3d4l$ ($l=1,3$),
$2s2p^63l^24l^{\prime\, 2}$ ($l=0,1,2; l^{\prime}=0,1,2,3$),
$2s2p^63l^24l^{\prime}4l^{\prime\prime}$
($l=0,1,2; l^{\prime}=0,1; l^{\prime\prime}=2,3$),
$2s2p^63s3d4l^2$ ($l=0,1,2,3$),
$2s2p^63l3l^{\prime}4l^{\prime\prime}4l^{\prime\prime\prime}$
($l=0,1; l^{\prime}=1,2; l^{\prime\prime}=0,1,2; l^{\prime\prime\prime}=1,2,3$).
We consider the same $LS$-terms as in the transitions from the $2p$ shell. 
Consequently, the numbers of CSFs included in the CI wavefunction expansion 
are 1695, 2865 and 3270  for the $J=1/2$, $3/2$, $5/2$ levels, respectively.

While constructing the upper-state CI wavefunction expansion configuration 
set for the $2s - 5l$ transition array, we apply the same principles as in the 
$n=4$ case but replace the configuration sets containing $4l$ electrons with 
those containing $5l$ electrons.  That expansion is extended with a set of 
even-parity configurations with a valence $n=4$ electron
$2p^63l^24l^{\prime}$ ($l=0,1,2; l^{\prime}=0,2$) 
and
$2p^63l3l^{\prime}4l^{\prime\prime}$
($l=0,1; l^{\prime}=1,2; l^{\prime\prime}=0,1,2,3$).
This set generates 1751, 2954 and 3364 CSFs in the CI wavefunction expansion 
for the $J=1/2$, $3/2$ and $5/2$  levels, respectively. 

\subsubsection{Transition data}
\label{td13}

Using the above methods to generate the CI wavefunction expansions, 
we obtained wavefunctions to produce photoexcitation line data for the 
Fe\,{\sc xiv} ion. Calculated line data cover the wavelength region 
$\lambda= 10.571 - 14.432$\,\AA \,\, for $2s - 3l$ transitions (a total of 617 
lines), $\lambda= 9.298 - 11.707$\,\AA \,\, for $2s - 4l$ transitions 
(3194 lines) and $\lambda= 8.890 - 10.861$\,\AA \,\, for $2s - 5l$ transitions 
(147 lines). For transitions from the $2p$ shell, the wavelength range
was $\lambda= 12.341 - 16.626$\,\AA \,\, for  excitation from the $n=3$ shell 
(1343 lines), $\lambda= 10.141 - 13.642$\,\AA \,\, for the $n=4$ shell 
(7425 lines), and $\lambda= 9.685 - 12.438$\,\AA \,\, for the $n=5$ shell
(8640 lines). 

Table~\ref{tab_fe13short} lists the atomic data for the Fe\,{\sc xiv} UTA lines. 
Transition wavelengths $\lambda$ and absorption oscillator 
strengths $f_{ij}$ for excitation from the $2s$ and $2p$ shells to the outer $n=3,4,5$
shells are given for lines with $f \geq 0.1$. A more extensive data set 
is presented in the
online version of this table, where we provide results for all lines with 
$f \geq 0.0001$ (see Sect.~\ref{trim} for more details about data trimming).

%%%%%%%%%%%% T3 %%%%%%%%%%%%%%%%%%%%%%%%%%%%%%%%%%
\begin{deluxetable}{ccccrr}
\tabletypesize{\scriptsize}
\tablecolumns{6} 
\tablewidth{0pt}
%\tablenum{1}
\tablecaption{
Transitions $2p - n_jd$ and $2s - n_jp $ from the $L$ shell to the $n_j = 3, 4, 5$ 
shells in Fe$^{13+}$. Column $n_il_i$ denotes the initial shell, $n_j$ the 
final shell for optical electron transition, $g_i$ and $g_j$ the statistical 
weights for the initial and final levels, $\lambda$ the line wavelength (in \AA)
and $f_{ij}$ the absorption oscillator strength.
\label{tab_fe13short}
} 
\tablehead{ 
\colhead{$n_il_i$} &
\colhead{$n_j$} & 
\colhead{g$_i$} &
\colhead{g$_j$} & 
\colhead{$\lambda $(\AA)} &
\colhead{$f_{ij}$}
}
\startdata 
2p& 3& 2&  4&  15.474&  0.920\\
2p& 3& 4&  6&  15.402&  0.430\\
2p& 3& 2&  2&  15.433&  0.394\\
2p& 3& 4&  6&  15.507&  0.393\\
2p& 3& 4&  4&  15.449&  0.378\\
2p& 3& 4&  2&  15.408&  0.310\\
2p& 3& 4&  4&  15.375&  0.266\\
2p& 3& 4&  6&  15.645&  0.209\\
2p& 3& 2&  2&  15.310&  0.201\\
2p& 3& 2&  4&  15.499&  0.194\\
2p& 3& 4&  4&  15.665&  0.164\\
2s& 3& 2&  4&  14.282&  0.157\\
2s& 3& 2&  2&  14.300&  0.147\\ 
2p& 3& 2&  4&  15.522&  0.147\\
2p& 4& 4&  6&  12.891&  0.159\\
2p& 4& 2&  4&  12.921&  0.109\\
2p& 3& 2&  2&  15.697&  0.103\\
\enddata 
\tablecomments {Table \ref{tab_fe13short} is published in its entirety in the 
electronic edition of the {\it Astrophysical Journal}. Only a portion is 
shown here for guidance regarding its form and content.}
\end{deluxetable} 
%%%%%%%%%%%%%%%%%%%%%%%%%%%%%%%%%%%%%%%%%%%%%%%%

\begin{figure}[!t]
%\plottwo{fe15.bw.eps}{fe15.eps}
%\plotone{fe13.bw.eps}
%\plotone{fe13.eps}
\includegraphics[scale=0.65]{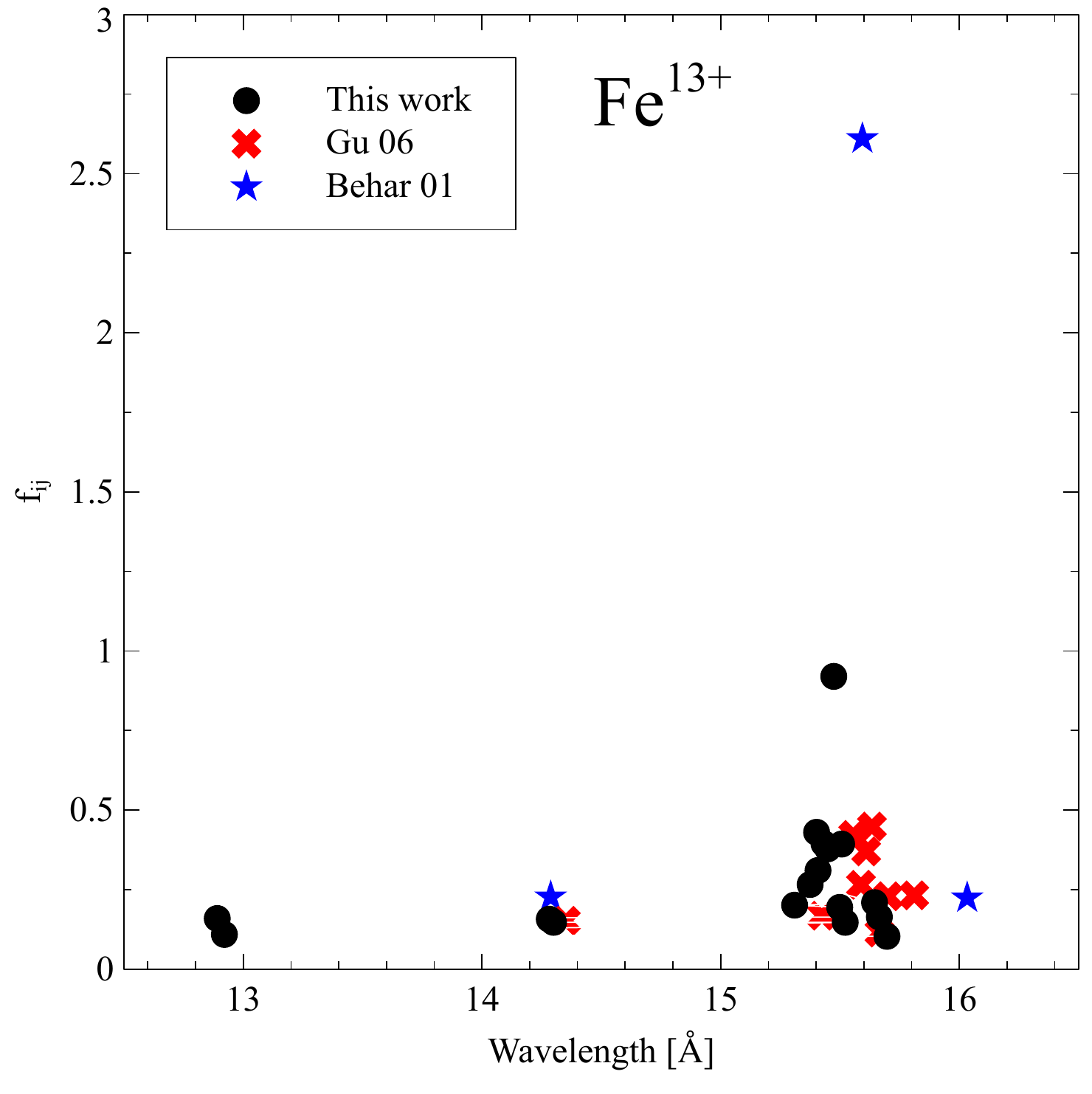}
\caption{A comparison of the wavelengths $\lambda$ and absorption oscillator
strengths $f_{ij}$ for the strongest lines in Fe\,{\sc xiv}.
Dots represent our data, crosses are the results of 
\citet{Gu2006} and stars those from \citet{Behar2001}. 
The data at $\lambda=15.5$\AA\, represent the $2p - 3d$ UTA, 
those at $\lambda=14.3$\AA\, represent the $2s - 3p$ UTA, and those at $\lambda=12.9$\AA\,
represent the $2p - 4d$ UTA.
The \citet{Behar2001} data are given for an $f$-value averaged wavelength 
$\lambda_{\mathrm{av}}$ and the total oscillator strength $\sum f_{ij}$ 
of the line group.
See the electronic edition of the Journal for a color version 
of this figure.\label{fig13}}
\end{figure}

In Fig. \ref{fig13} we compare the strongest lines from our calculation with
the results derived using the MBPT approach of \citet{Gu2006} and with the 
HULLAC  relativistic data from \citet{Behar2001}.  There is reasonable 
wavelength agreement for all three sets, as found
for Fe\,{\sc xvi} and Fe\,{\sc xv}. The only exception is the line
with $\lambda = 16.033$\AA\ from \citet{Behar2001}, with the larger values of 
oscillator strengths $\sum f_{ij}$ found by these authors
explained in Sect.~\ref{td15}. A wavelength
disagreement between our results and the MBPT data was also discussed in
that section.

%%%%%%%%%%%%%%%  F3  %%%%%%%%%%%%%%%%%%%%%%%%%%%%%%%%%

\clearpage

\section{Incorporation into Cloudy}

\subsection{Data sources}
\label{sources}

The UTA data sets now used in Cloudy are summarized in  
Table \ref{tab_data_sources}.  
The Opacity Project (OP; \citealp{BadnellEtAl05}) produced complete data,
for the $K$ shell as well as the $L1$ and $L2$  shells where appropriate,
for ions with twelve or fewer bound electrons.  
These include the auto-ionization branching ratio which we consider 
in calculating the ionization rate.
The OP data are marked ``B'' in this table.  
\citet{Gu2006} provide all stages of ionization for Fe but includes only excitations from 
the $2p$ shell, 
denoted by ``G'' in the Table.  
These data sets have been included in Cloudy since shortly after their original
publication dates.

The data presented in this paper are now being incorporated into
Cloudy and are denoted by ``K'' in Table \ref{tab_data_sources}.  
Two ions, Fe$^{14+}$ and Fe$^{15+}$, are present in both 
the current data and the OP.  
We use the OP data for excitations from the $K$ shell and the current results
for $L$ shell excitations.  
Our new data do not include autoionization rates, so these were
copied from \citet{Gu2006} for the 5 most important lines of Fe\,{\sc XV} and
Fe\,{\sc XVI}.
For other lines we assume that the branching ratio for 
ionization is 100\%. In the default setup of the code, we only use
transitions from the ground level for all data sources.
See also the discussion in \S~\ref{density_indicator}.

It is clear from Table \ref{tab_data_sources} that many ions simply
have no current data.  
For example, there are no $K$ shell data for Fe$^{13+}$.
These are all major requirements for accurate spectral simulations.

%%%%%%%%%%%%%%%  T3  %%%%%%%%%%%%%%%%%%%%%%%%%%%%%%%%%
\begin{deluxetable}{lrrrrrrrrrrrrrrccrrrrrrrrrrrrrrrrr}
\tabletypesize{\scriptsize}
\tablecolumns{32} 
\tablewidth{0pt}
\tabcolsep 4pt
%\tablenum{1}
\rotate
\tablecaption{
Data sources  used in the default setup of the  the Cloudy code. B -- Badnell et al. (2005) data, G -- Gu et al. (2006) data,
K -- present results.
\label{tab_range}
\label{tab_data_sources}
} 
\tablehead{ 
\colhead{Ion} &
\colhead{0} & 
\colhead{1} & 
\colhead{2} & 
\colhead{3} & 
\colhead{4} & 
\colhead{5} & 
\colhead{6} & 
\colhead{7} & 
\colhead{8} & 
\colhead{9} & 
\colhead{10} & 
\colhead{11} & 
\colhead{12} & 
\colhead{13} & 
\colhead{14} & 
\colhead{15} & 
\colhead{16} & 
\colhead{17} & 
\colhead{18} & 
\colhead{19} & 
\colhead{20} & 
\colhead{21} & 
\colhead{22} & 
\colhead{23} & 
\colhead{24} & 
\colhead{25} & 
\colhead{26} & 
\colhead{27} & 
\colhead{28} & 
\colhead{29} & 
\colhead{30} & 
}
\startdata 
H&&&&&&&&&&&&&&&&&&&&&&&&&&\\
He&&&&&&&&&&&&&&&&&&&&&&&&&&\\
Li&B&&&&&&&&&&&&&&&&&&&&&&&&&\\
Be&B&B&&&&&&&&&&&&&&&&&&&&&&&&\\
B&B&B&B&&&&&&&&&&&&&&&&&&&&&&&\\
C&B&B&B&B&&&&&&&&&&&&&&&&&&&&&&\\
N&B&B&B&B&B&&&&&&&&&&&&&&&&&&&&&\\
O&B&B&B&B&B&B&&&&&&&&&&&&&&&&&&&&\\
F&B&B&B&B&B&B&B&&&&&&&&&&&&&&&&&&&\\
Ne&B&B&B&B&B&B&B&B&&&&&&&&&&&&&&&&&&\\
Na&B&B&B&B&B&B&B&B&B&&&&&&&&&&&&&&&&&\\
Mg&B&B&B&B&B&B&B&B&B&B&&&&&&&&&&&&&&&&\\
Al&&B&B&B&B&B&B&B&B&B&B&&&&&&&&&&&&&&&\\
Si&&&B&B&B&B&B&B&B&B&B&B&&&&&&&&&&&&&&\\
P&&&&B&B&B&B&B&B&B&B&B&B&&&&&&&&&&&&&\\
S&&&&&B&B&B&B&B&B&B&B&B&B&&&&&&&&&&&&\\
Cl&&&&&&B&B&B&B&B&B&B&B&B&B&&&&&&&&&&&\\
Ar&&&&&&&B&B&B&B&B&B&B&B&B&B&&&&&&&&&&\\
K&&&&&&&&B&B&B&B&B&B&B&B&B&B&&&&&&&&&\\
Ca&&&&&&&&&B&B&B&B&B&B&B&B&B&B&&&&&&&&\\
Sc&&&&&&&&&&B&B&B&B&B&B&B&B&B&B&&&&&&&\\
Ti&&&&&&&&&&&B&B&B&B&B&B&B&B&B&B&&&&&&\\
V&&&&&&&&&&&&B&B&B&B&B&B&B&B&B&B&&&&&\\
Cr&&&&&&&&&&&&&B&B&B&B&B&B&B&B&B&B&&&&\\
Mn&&&&&&&&&&&&&&B&B&B&B&B&B&B&B&B&B&&&\\
Fe&G&G&G&G&G&G&G&G&G&G&G&G&G&K&BK&BK&B&B&B&B&B&B&B&B&&\\
Co&&&&&&&&&&&&&&&&B&B&B&B&B&B&B&B&B&B&\\
Ni&&&&&&&&&&&&&&&&&B&B&B&B&B&B&B&B&B&B\\
Cu&&&&&&&&&&&&&&&&&&B&B&B&B&B&B&B&B&B&B\\
Zn&&&&&&&&&&&&&&&&&&&B&B&B&B&B&B&B&B&B&B\\
\tableline
Ion&0&1&2&3&4&5&6&7&8&9&10&11&12&13&14&15&16&17&18&19&20&21&22&23&24&25&26&27&28 &29&30\\
\enddata 
\end{deluxetable} 

%%%%%%%%%%%%%%%%%%%%%%%%%%%%%%%%%%%%%%%%%%%%%%%%

\clearpage

\subsection{Data trimming}
\label{trim}
% atmdat_readin variable f_cutoff determines cutoff

\begin{figure}[!ht]
%\plottwo{fe15.bw.eps}{fe15.eps}
%\plotone{fe13.bw.eps}
%\plotone{Fe_13_trimming.eps}
\includegraphics[scale=0.65]{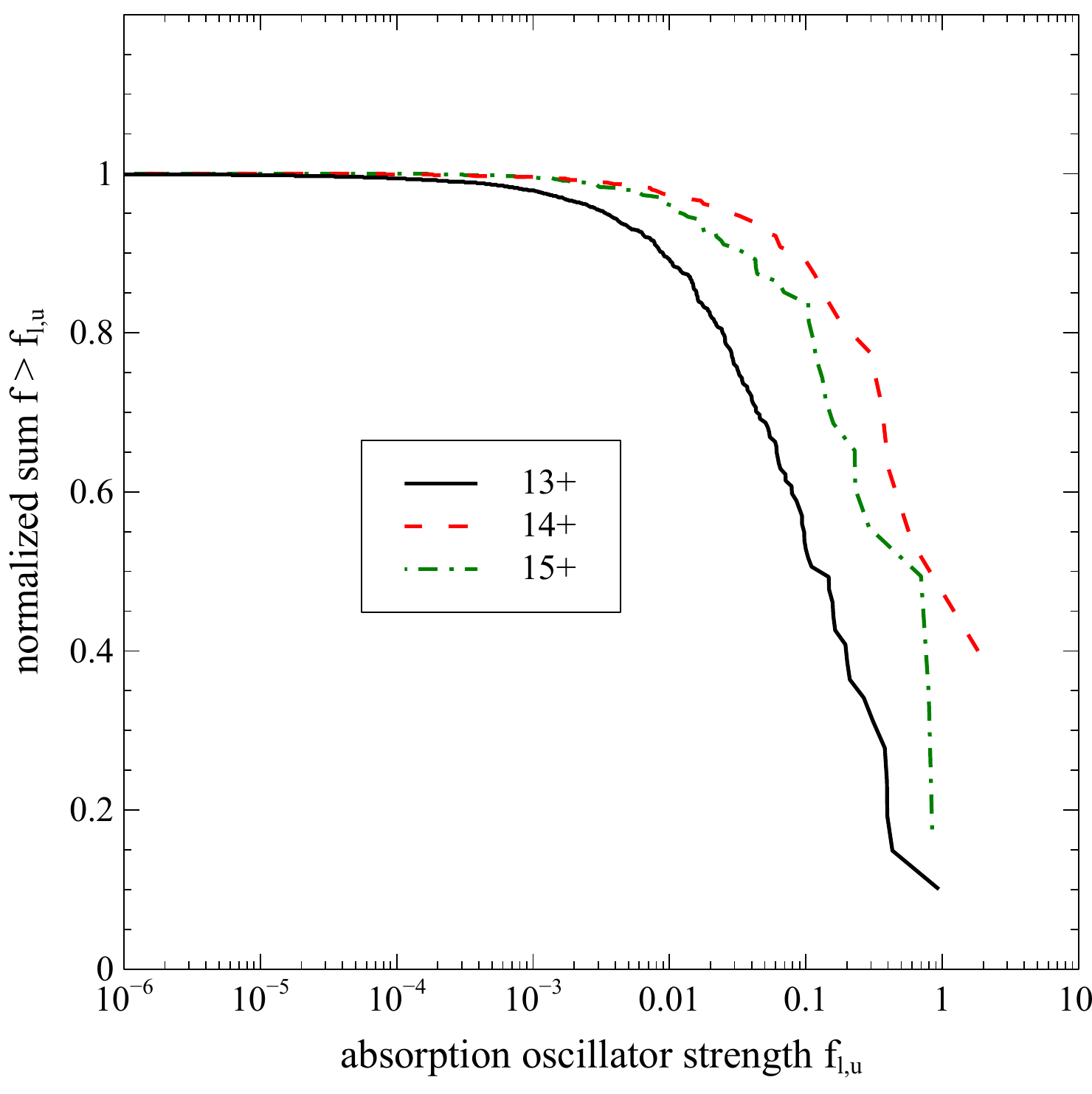}
\caption{Absorption line oscillator strength $f_{l,u}$ plotted against 
the sum of all lines with $f > f_{l,u}$.  
The y-axis has been scaled by the sum of the oscillator strengths.
The integrated absorption is well converged when only lines with
$f > 10^{-4}$ are included.
\label{fig:FeTrimming}}
\end{figure}

The newly computed data sets are quite extensive 
and include a large number of exceptionally weak lines.  
Figure \ref {fig:FeTrimming} shows the absorption line oscillator 
strength $f_{l,u}$ plotted against 
the normalized sum of all lines with $f > f_{l,u}$.  
Careful examination of such data shows that the summed oscillator strength, 
which affects the UTA ionization rate, 
has converged to within 0.57\%, 0.15\%, and 0.08\% of the total
for Fe$^{13+}$, Fe$^{14+}$, and Fe$^{15+}$,
when all lines with $f \geq 10^{-4}$ are included.  
Weaker lines have little effect upon the ionization rate because 
the line-center opacity is smaller than the continuous photoionization opacity.
Accordingly, we only consider lines with $f \geq 10^{-4}$, although
we retain all data for future flexibility.
For the case of Fe$^{13+}$ this reduces the number of lines 
from 80894 to 754, for Fe$^{14+}$ from 2738 to 103,
and for Fe$^{15+}$ from 595 to 137.

%%%%%%%%%%%%%%%  F4  %%%%%%%%%%%%%%%%%%%%%%%%%%%%%%%%%
%%%%%%%%%%%%%%%%%%%%%%%%%%%%%%%%%%%%%%%%%%%%%%%%

\subsection{Damping parameters and line broadening}

The Voigt function, which describes the line profile including both 
thermal and natural broadening, is given by
\begin{equation}
H\left( {a,x} \right) = \frac{a}{\pi }\int_{ - \infty }^{ + \infty } 
{\frac{{\exp \left( { - {y^2}} \right)}}{{{{\left( {x - y} \right)}^2} + {a^2}}}} \,dy
\end{equation}

\noindent
where $x$ is the displacement from line center, measured in terms of the
Doppler width $\Delta \nu _{Dop}$,
\begin{equation}
x \equiv \frac{{\nu  - {\nu _0}}}{{\Delta {\nu _{Dop}}}}  .
\end{equation}
\noindent
The damping parameter $a$ is the ratio of 
natural (radiation damped) to thermal line widths
\begin{equation}
a \equiv \frac{\gamma }{{4\pi \Delta {\nu _{Dop}}}},
\end{equation}
\noindent
and the natural broadening width is given by the damping constant $\gamma$ which
is expressed as a sum of radiative $A^r$ and autoionization $A^a$ rates:
\begin{equation}
\gamma  = \sum\limits_{l < u} {{A_{ul}^r}} + 
\sum\limits_{l^{\prime} < u} {{A_{ul^{\prime}}^a}} 
\end{equation}
$H(a,x)$ is normalized so that its integral over $x$ is $\sqrt{\pi}$.

The Voigt function must be evaluated to account for line
self-shielding when finite column densities are encountered, while 
the total radiative decay ratio out of the upper level is needed
to derive the damping constant $\gamma$.
This lifetime is the sum of the autoionization and radiative decay rates.
Our new data do not include calculations of auto ionization rates (except for the 5 lines mentioned earlier).
However, in order to calculate the damping constant $\gamma$ we need to assume a value, so
we adopted the value, $10^{15} \ps$, 
taken from Opacity Project results for Fe~{\sc xv} and Fe~{\sc xvi}.
The other data sources include auto ionization rates, which we adopt.

\begin{figure}
  \centering
\includegraphics[scale=0.5]{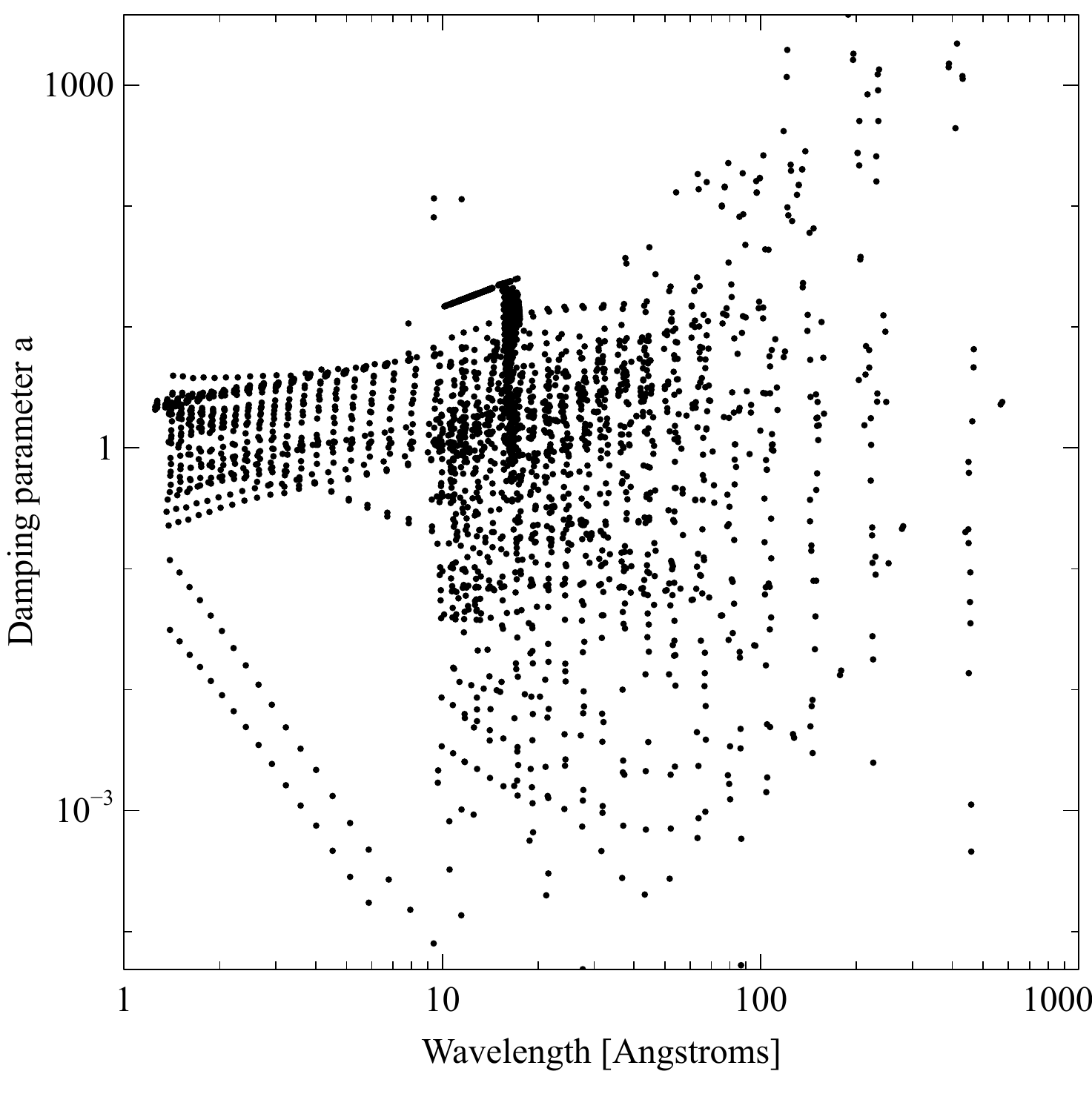}
  \caption{Damping parameters as a function of the line
  wavelength for our complete set of UTA lines.}
  \label{fig:WLvsDamp}
\end{figure}

Figure \ref{fig:WLvsDamp} shows the resulting damping parameters,
plotted as a function of transition wavelength, for a gas kinetic temperature of $10^4$~K.
UTA transitions are strongly damped, often with $a \gg 1$,
because of the rapid autoionization rate.
For comparison, the damping parameter for a strong UV line, such as 
\ion{H}{1} L$\alpha$, is $a \sim 4\times 10^{-4}$.
It should be noted that the high damping parameter $a$ is a consequence of 
the inner-shell nature of these excitations regardless of their being in UTAs.

%%%%%%%%%%%%%%%  F5  %%%%%%%%%%%%%%%%%%%%%%%%%%%%%%%%%

%%%%%%%%%%%%%%%%%%%%%%%%%%%%%%%%%%%%%%%%%%%%%%%%

While simple approximations to the Voigt function may be sufficient
for UV -- optical spectroscopy, with small $a$, the large range in
damping parameter that occurs when UTAs are considered make it
important to use implementations of the Voigt function which are
accurate for all $a$.  We adopt the routine provided by
\citet{Wells99}, combined with a specially designed faster routine for
$a \leq 0.1$, which provides results with a relative accuracy of 1 in
$10^4$, and has been confirmed to pass all the test cases given by
\citet{Zaghloul11}.

\begin{figure}
  \centering
\includegraphics[scale=0.5]{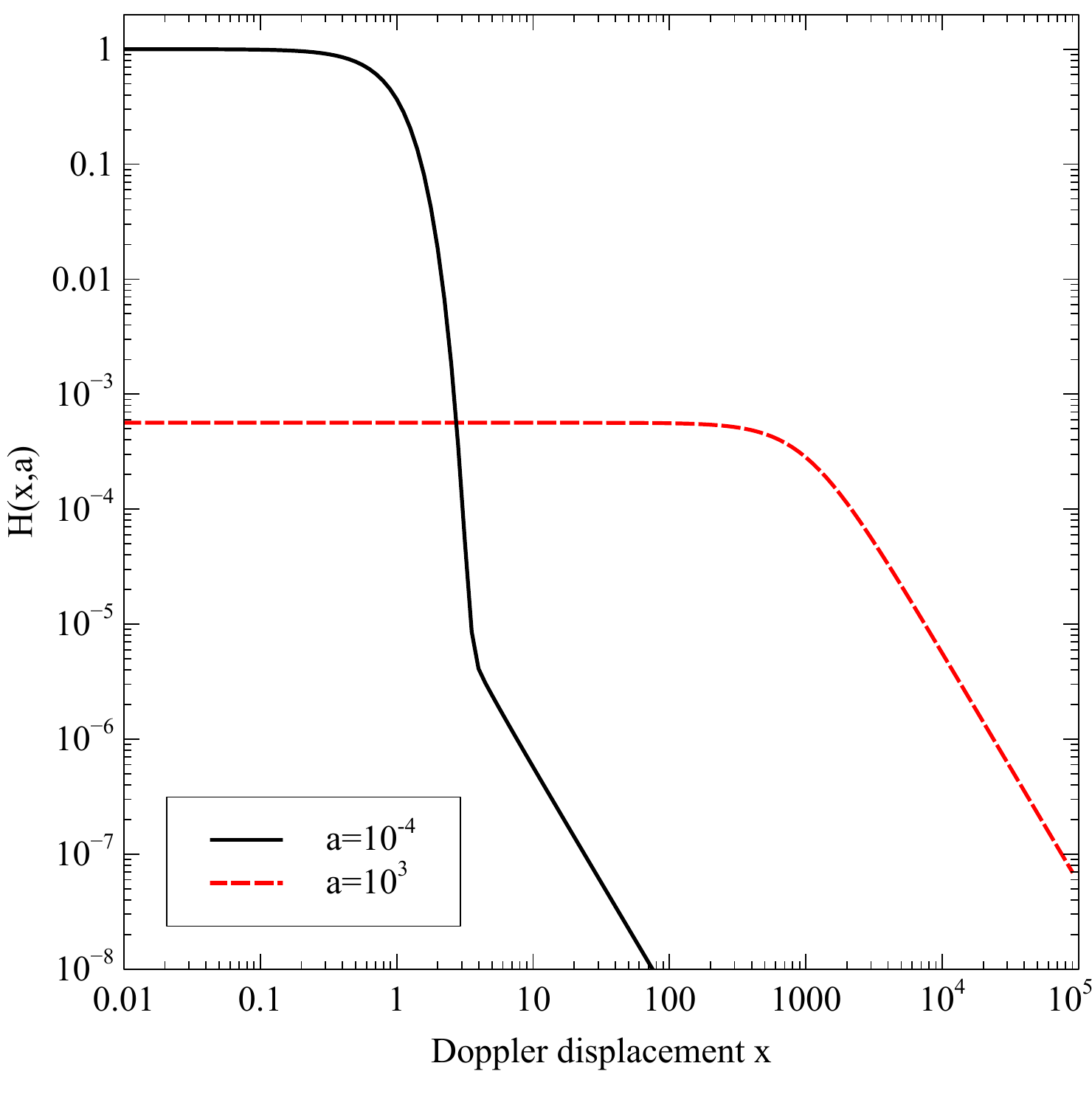}
  \caption{The Voigt function $H(a,x)$ for 
  two values of the damping parameter $a$.}
  \label{fig:voigt}
\end{figure}

Stellar atmosphere texts \citep{Rutten2002,Mihalas1978} often focus on
results derived for approximations to the Voigt function which, while
valid in the original context, are not accurate for the full range of
frequency and $a$ needed to include UTA transitions.  Figure
\ref{fig:voigt} shows the function for a typical strong UV line such
as \ion{H}{1} Ly$\alpha$ ($a \sim 10^{-4}$) and a strongly damped UTA
transition ($a \sim 10^3$).  We see that at line center $H\left( {a,x}
\right) \sim \left( {1 + a} \right)^{ - 1}$, and that the core of the
line is roughly $x \sim \left( {1 + a} \right)$ wide.

%%%%%%%%%%%%%%%  F6  %%%%%%%%%%%%%%%%%%%%%%%%%%%%%%%%%
%%%%%%%%%%%%%%%]%%%%%%%%%%%%%%%%%%%%%%%%%%%%%%%%%

There are a large number of lines which strongly overlap to
produce the observed UTA features.
The effects of mutual line shielding due to overlap must be included
if the ionization rate is to be properly computed.
Line overlap is treated using a combination of a coarse, low resolution
continuum, and a fine continuum, with resolution sufficient to resolve lines,
as described in \citet{ShawEtal05}.

\section{Application to AGN}

\subsection{Fe XIV as a density indicator}
\label{density_indicator}

Our \ion{Fe}{14}\ data include
both fine-structure levels within the ground term, namely $^2P_{1/2}^{\mathrm o}$ 
and  $^2P_{3/2}^{\mathrm o}$. These are separated by 2.33741 eV and produce 
the famous ``green'' Fe\,{\sc xiv} coronal line at 5303\,\AA . There are 
approximately the same number of UTA lines originating from either level in 
the overall configuration.

The critical density  of the excited $J=3/2$ level is $\sim 3\times 10^9$ 
cm$^{-3}$ at 10$^5$~K. The temperature $T = 10^5$~K is approximately that of
the formation of the Fe ions discussed in current paper. At considerably lower 
densities most of the population will be in the $J=1/2$ level, while for high 
temperatures and densities the levels will be populated according to their 
statistical weight and most will be in the excited level. If the absorption 
characteristics are different in these two different limits then the UTA lines 
could be used as a density indicator.

%%%%%%%%%%%%%%%  F7  %%%%%%%%%%%%%%%%%%%%%%%%%%%%%%%%%
\begin{figure}[!ht]
%created with Fe14_density2.vsz in same directory
\includegraphics[scale=0.55]{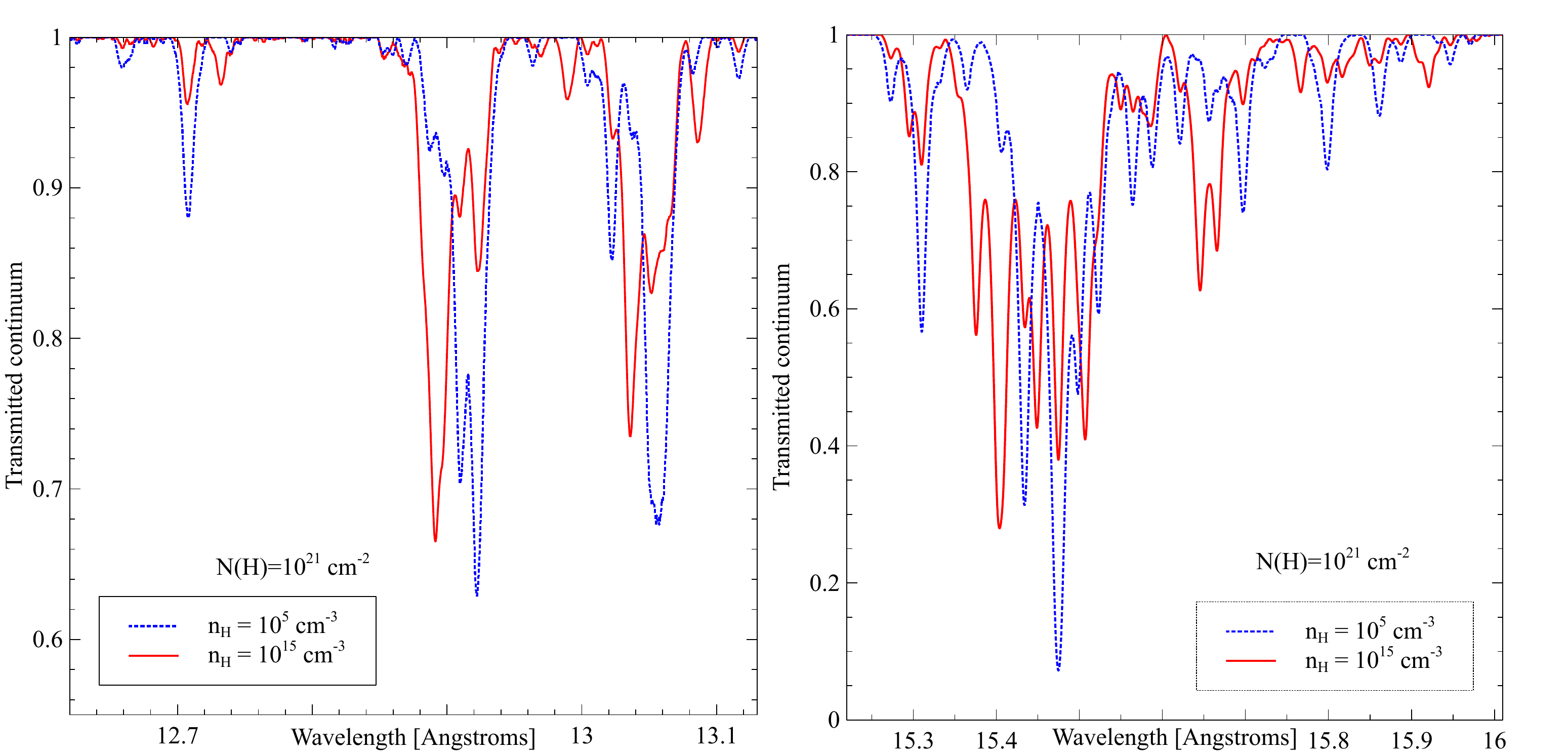}
\caption{UTA spectrum of \ion{Fe}{14} transmitted through a cloud with a column density
of N(H)$= 10^{21}$\,cm$^{-2}$ and a solar Fe abundance.
The effects of different densities upon the shape of the
\ion{Fe}{14} UTA are shown, with the low density case in blue having nearly all populations in the 
lower $J=1/2$ level, while the high density case has level populations within 
the ground term proportional to their statistical weight.
This calculation only included \ion{Fe}{14}.
\label{fig:Fe14density}}
\end{figure}
%%%%%%%%%%%%%%%%%%%%%%%%%%%%%%%%%%%%%%%%%%%%%%%%

Figure \ref{fig:Fe14density} (right frame) shows how the profile of the $\lambda 15.5\,$\AA\ 
\ion{Fe}{14} UTA changes with density. The calculation assumed a total hydrogen 
column density of $10^{21}$ cm$^{-2}$, a solar Fe abundance, and two hydrogen 
densities, $n_{\rm H} = 10^5$ cm$^{-3}$, sufficiently low for all populations to lie
in the lower $J=1/2$ level, and $n_{\rm H} = 10^{15}$ cm$^{-3}$, high enough for
the levels to be populated according to their statistical weights. Significant 
differences are present. The features originating from the configurations which
include the excited $J=3/2$ level
are not present in the low density profile. At the high density limit, the ratio of
the excited and lower level population is 2:1. Therefore, new features connected 
to the $J=3/2$ level (e.g. $\lambda 15.35\,$\AA, $\lambda 15.40\,$\AA,
$\lambda 15.65\,$\AA) become evident, whereas features originating from the 
ground $J=1/2$ level (e.g. $\lambda 15.30\,$\AA, $\lambda 15.43\,$\AA,
$\lambda 15.47\,$\AA) become less pronounced. 
The total
absorption does not change because the sum of the absorption 
oscillator strengths in this wavelength region is approximately the same 
both for $J=3/2$ and $J=1/2$ levels, although different transitions are
produced by each level.

Figure \ref{fig:Fe14density} (left frame) shows the effects of density in the area of the
$\lambda 13.0\,$\AA\ UTA lines in \ion{Fe}{14}. It demonstrates very similar 
behavior to the case of the $\lambda 15.5\,$\AA\ UTA shown in the right panel,
with
$J=3/2$ lines appearing at higher densities and $J=1/2$ lines being prominent
at low-density limit.

Future generations of X-ray spectrometers may be able to use such differences 
to measure the density of the absorbing gas. This would make it possible to 
deduce the location of the gas. Nevertheless, improved accuracy of the 
theoretical wavelengths would be needed for such a test to be definitive.

Absorption line spectroscopy generally has a degeneracy introduced by
the fact that the ionization distribution determines 
an ionization parameter, the ratio of the flux of ionizing photons to gas 
density. 
Literature on this type of analysis is extensive, and summarized by
\citet{Chakravorty2009}.
The ionization itself is not sensitive to either flux or density.
By measuring the density and ionization parameter, the flux could be 
deduced, which then leads to the determination of the source - cloud separation.

Tests show that the density effects do not have a significant effect on the 
ionization of the gas because the total absorption oscillator strengths are similar. 
In the current implementation we assume the low density 
limit in computing the effects of UTA transitions and the resulting spectra.

% The above calculation was done as one-off hack of branch/romas3
% hacked source file is committed in models / Fe14_density, must
% set lgAllowSplitFe14 true and include "fudge 0" in input stream to
% assume pops are distributed according to 2J+1 at high density,
% otherwise all in ground level for low density

The ground terms of Fe$^{14+}$ and Fe$^{15+}$ are simpler,
having only one level,
so such distinctions do not occur.

\subsection{Probing the thermal stability of the warm absorber}

The nature of the absorbing and emitting clouds in AGN is a long-standing
problem \citep{AGN3,Chakravorty2009}.
By analogy with the local ISM, several gas phases are thought to be 
present at one location, each in equilibrium with the radiation field, but
having different levels of ionization and a range of temperatures, but
the same gas pressure.
This has been discussed extensively in the literature
\citep{1981ApJ...249..422K,
1997ApJ...478...94H,
2001ApJ...561..684K,
2001A&A...374..914K,
1995MNRAS.273.1167R}.

\begin{figure}[!ht]
%\plottwo{fe15.bw.eps}{fe15.eps}
%\plotone{fe13.bw.eps}
%\plotone{Fe_13_trimming.eps}
\includegraphics[scale=0.65]{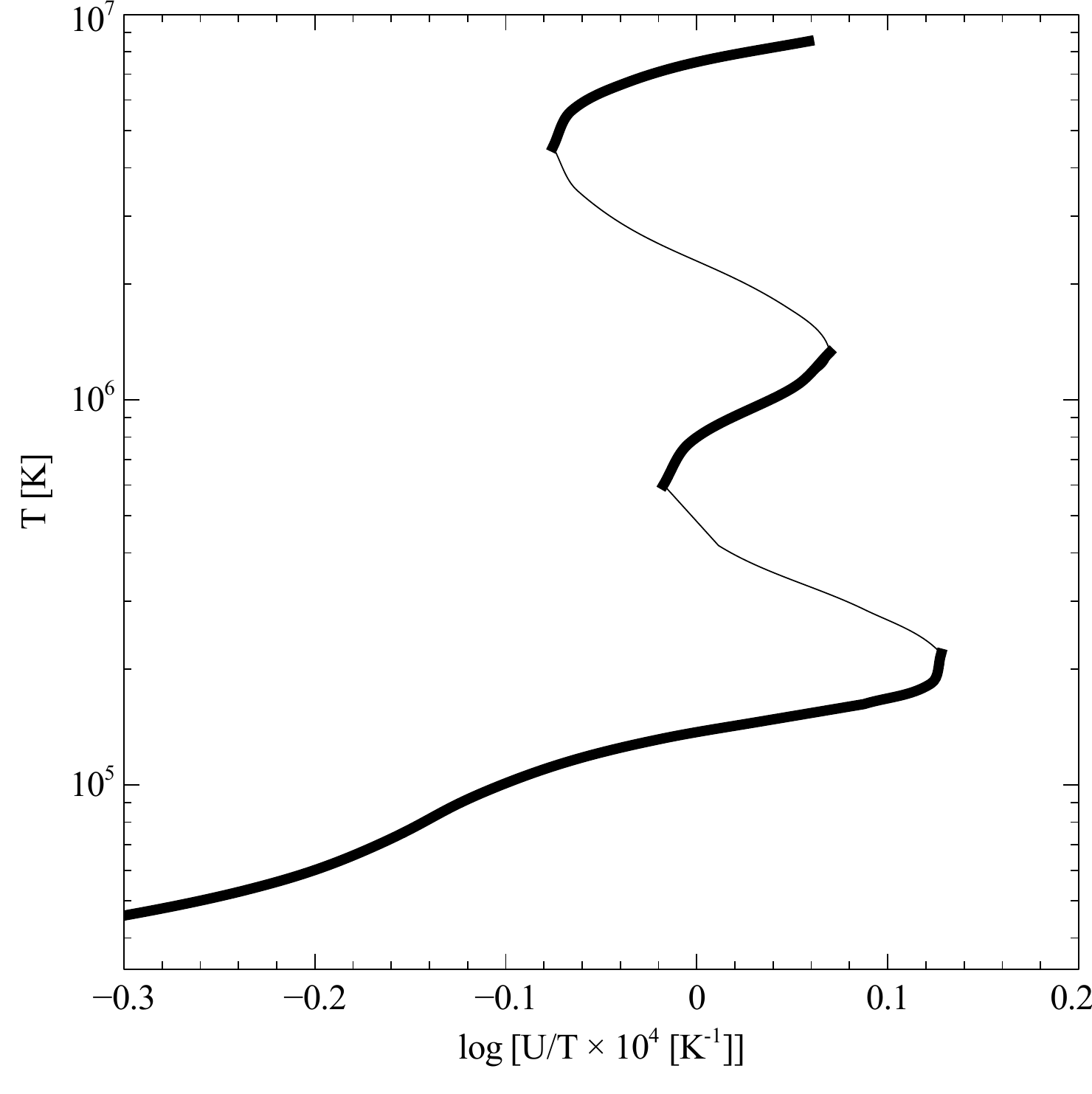}
\caption{A newly-computed AGN thermal stability curve.
The default AGN SED described by \citet{Chakravorty2009}
and a hydrogen density of $10^8 \pcc$ are assumed.
Thermally stable regions are shown by the thicker lines.
\label{fig:stability}}
\end{figure}

The ionization, temperature and spectrum of clouds that can exist
is determined by the type of stability analysis shown in Figure \ref{fig:stability}.
This shows the familiar ``S curve'', which is computed for an optically
thin cell of gas exposed to the AGN radiation field.
Thermally stable phases have positive slope, while regions with
negative slope are unstable so that gas will only exist in these regions
for a short time.
The shape of the S curve determines the properties of clouds which 
are thermally stable and so live long enough to contribute to the observed spectrum.

%%%%%%%%%%%%%%%  F8  %%%%%%%%%%%%%%%%%%%%%%%%%%%%%%%%%
%%%%%%%%%%%%%%%%%%%%%%%%%%%%%%%%%%%%%%%%%%%%%%%%

Recent S-curve calculations have focused on how properties of the
AGN might affect its shape, and so determine which clouds might exist.
The literature on this topic is vast, and recent examples include 
\citet{2007ApJ...663..799H},
\citet{2010ApJ...708..981H} and
\citet{2012ApJ...747...71H}, who examine observational determinations of the
stable points on the S curve,
\citet{Chakravorty2008} who show the effects
of updated dielectronic recombination rate coefficients,
\citet{Chakravorty2009},  who do a systematic stability analysis and and
\citet{Chakravorty2012},  who discuss the influence
of the SED on the cloud stability.

Here we show the effects of the improvements discussed in this paper, using
version C13 of Cloudy \citep{Ferland13}.
Some other recent improvements to the atomic data,
which focus on the atomic models used to compute the cooling,
are described by 
\citet{LykinsEtAl2013}.
Figure \ref{fig:stability} shows the thermal stability curve resulting from 
the improved atomic physics in this paper.
Many model parameters are taken from \citet{Chakravorty2009}.
The x-axis is the ratio of the dimensionless ionization 
parameter U, defined as the ratio of ionizing photon to hydrogen densities,
to the gas kinetic temperature, while the latter is the y-axis.
Thermally stable regions, those with positive slope, are shown
as the heavier lines.

In a series of papers
\citet{2007ApJ...663..799H},
\citet{2010ApJ...708..981H} and
\citet{2012ApJ...747...71H}  infer from observations of column
densities of certain ions that  temperatures between $4.5 < \textrm{log}\ T <  5.0$ are missing 
and ascribe it to thermal instability. On the other hand, this region
is stable in the current calculation with its assumed parameters.
The shape of the stability curve is affected by several other ingredients besides the atomic data. 
The composition assumed, and the form of the SED, also change it.
This suggests that the presence or absence of stable gas could be used to infer
the SED or gas composition, among other properties of the AGN.

\begin{figure}[!ht]
%\plottwo{fe15.bw.eps}{fe15.eps}
%\plotone{fe13.bw.eps}
%\plotone{Fe_13_trimming.eps}
\includegraphics[scale=0.65]{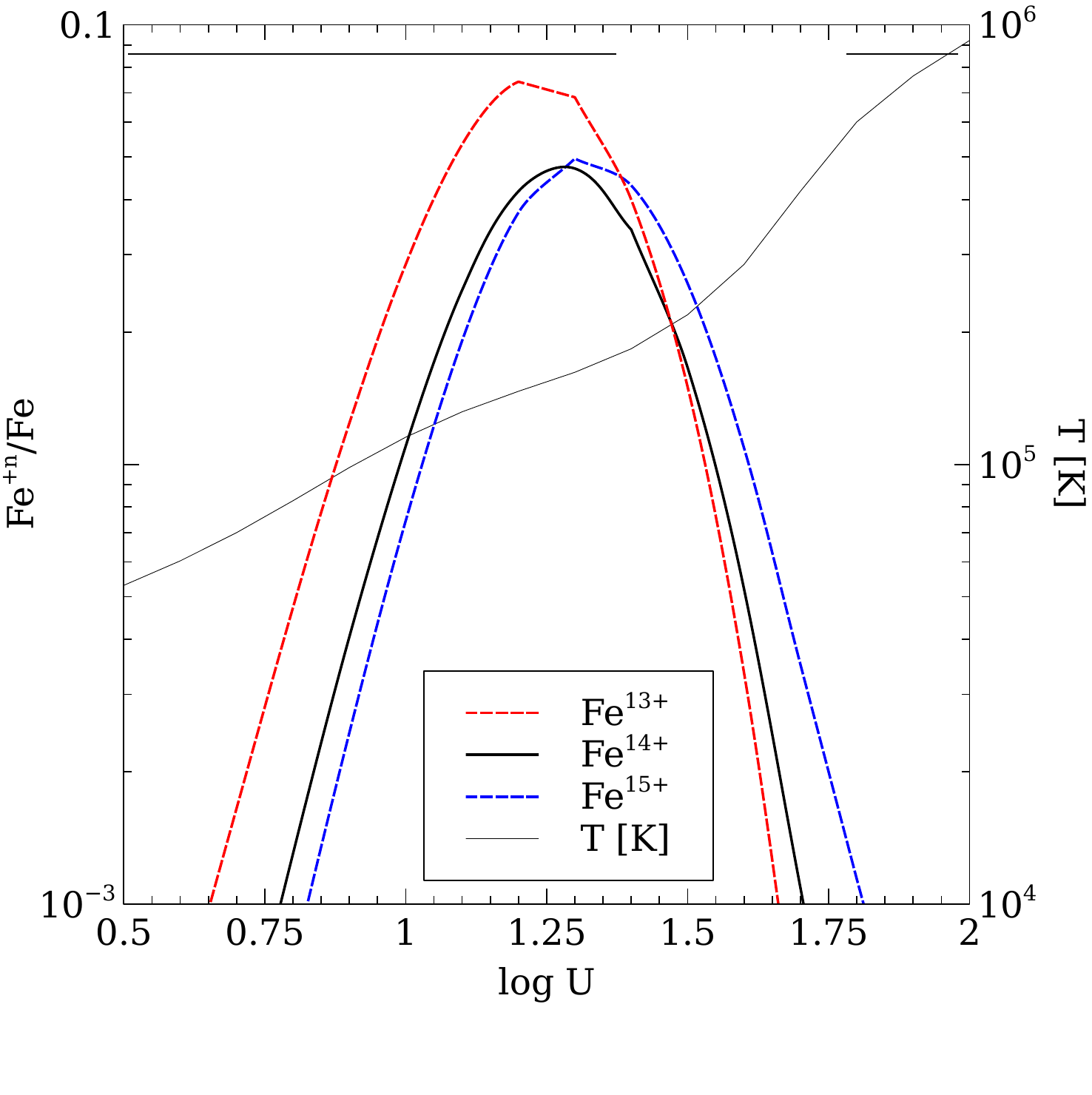}
\caption{The Fe ionization distribution across Figure \ref{fig:stability}, where 
the gas kinetic temperature is indicated by the solid line increasing from left
to right. 
Horizontal lines at the top of the figure mark the thermally stable 
regions of Figure \ref{fig:stability}.
The Fe ions considered in this paper probe the high temperature
end of the cool branch in Figure \ref{fig:stability},
unstable regions, and the low temperature end of the middle stable branch.
\label{fig:FeIonization}}
\end{figure}

Figure \ref{fig:FeIonization} shows the distribution of ionization stages of Fe
as a function of the ionization parameter U.
The ions discussed in this paper peak in the range log $U \sim 1 - 2$.
These probe the upper bound of the low-$T$
branch of thermally stable gas, the lowest unstable region, and the low-$T$ 
part of the middle stable branch.
Horizontal lines in the upper part of Figure \ref{fig:FeIonization}
indicate the regions where the gas is stable.

%%%%%%%%%%%%%%%  F9  %%%%%%%%%%%%%%%%%%%%%%%%%%%%%%%%%
%%%%%%%%%%%%%%%%%%%%%%%%%%%%%%%%%%%%%%%%%%%%%%%%

\begin{figure}[!ht]
%\plottwo{fe15.bw.eps}{fe15.eps}
%\plotone{fe13.bw.eps}
%\plotone{Fe_13_trimming.eps}
\includegraphics[scale=0.55]{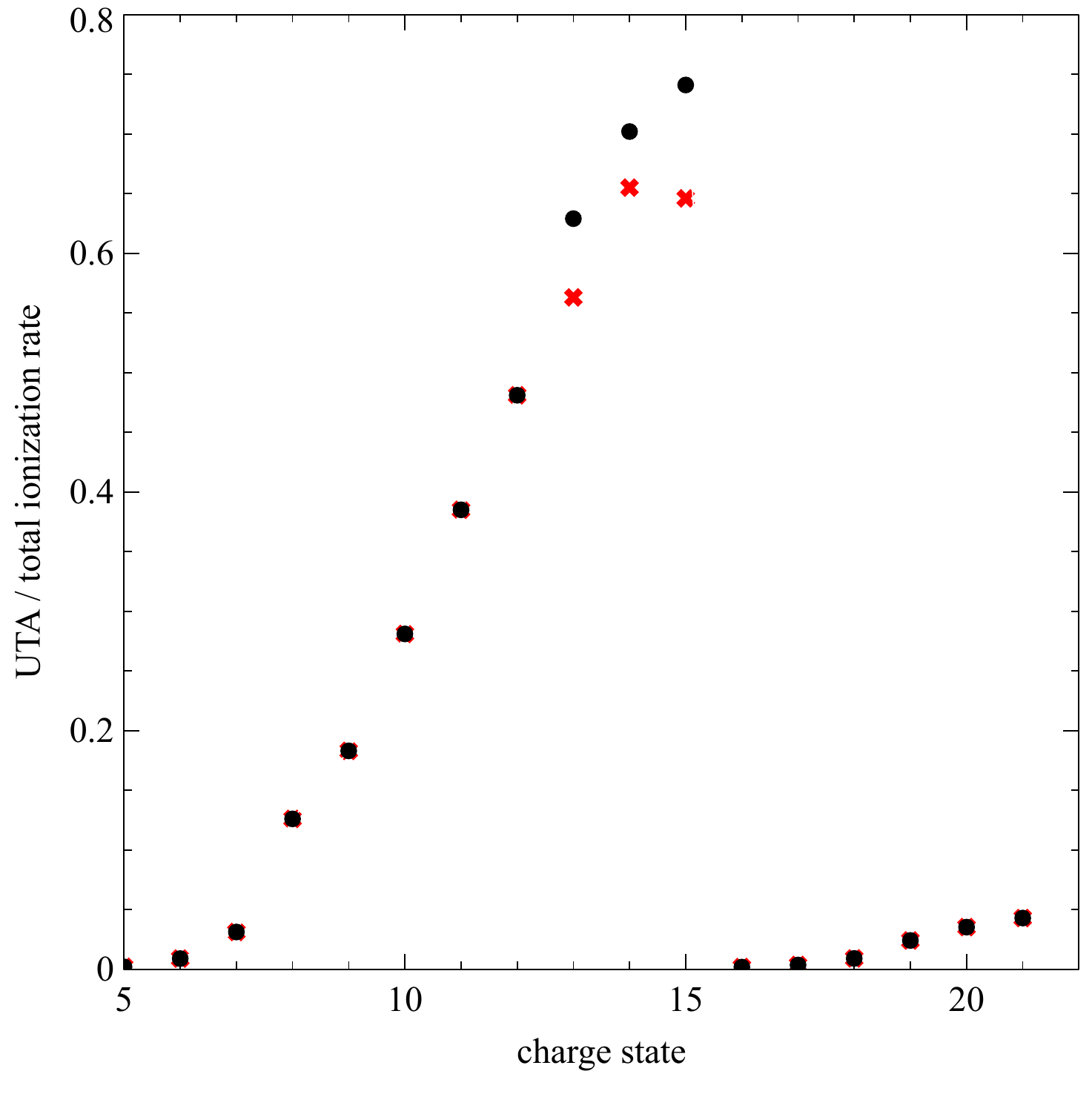}
\caption{The ratio of the UTA ionization to total ionization rate
for several Fe charge stages. Filled circles include
the results presented in this paper, while the crosses
represent the older data.
These are for the conditions occurring across Figure \ref{fig:stability}.
Rates are evaluated at the illuminated face so line self-shielding is not 
important.
UTA ionization is the dominant ionization process for the ions
considered in this paper.
\label{fig:UTAfraction}}
\end{figure}

The atomic data presented in this paper affect the details of
the transition between the cool and middle stable branches.
Figure \ref{fig:UTAfraction} shows the ratio of UTA to total 
ionization for several charge states of Fe, where the filled circles include
the data presented in this paper, while the crosses
represent the older data.
The larger number of lines derived here results in a larger UTA ionization 
rate, as shown in the Figure. By UTA ionization we mean the effect of 
autoionization following inner-shell photo-excitation. UTAs have the greatest 
effect on the ions which occur around the transition from the cool to middle 
stable branch, and their physics affects the details of this transition. 

%%%%%%%%%%%%%%%  F10  %%%%%%%%%%%%%%%%%%%%%%%%%%%%%%%%%
%%%%%%%%%%%%%%%%%%%%%%%%%%%%%%%%%%%%%%%%%%%%%%%%

\begin{figure}[!ht]
%\plottwo{fe15.bw.eps}{fe15.eps}
%\plotone{fe13.bw.eps}
%\plotone{Fe_13_trimming.eps}
\includegraphics[scale=0.85]{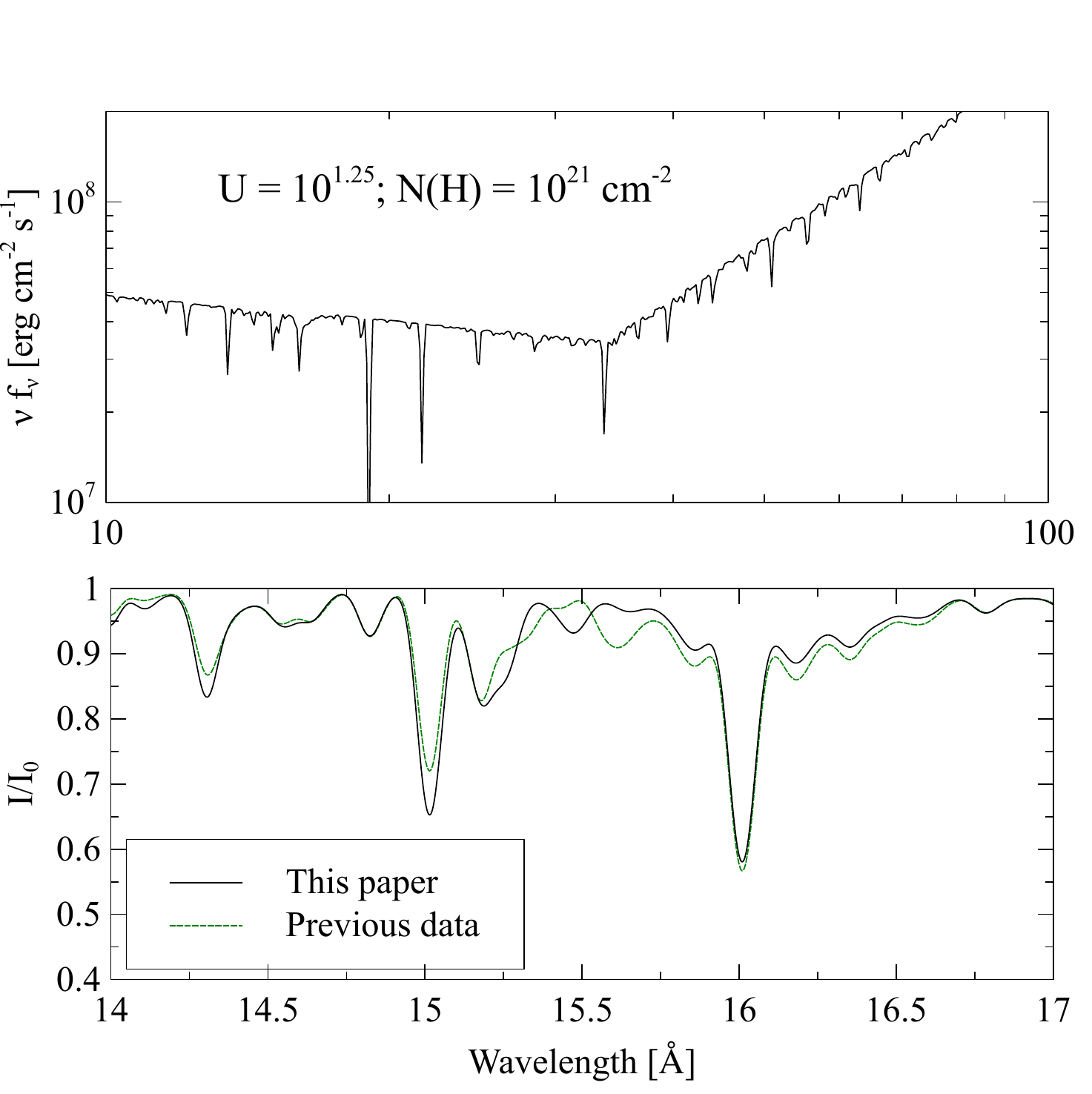}
\caption{The spectrum of a cloud lying along the line-of-sight to an AGN.
Cloud parameters are indicated in the upper panel, which shows
the transmitted coarse continuum.
The lower panel shows a small portion of the fine continuum with
predictions using the atomic data summarized in this paper in black,
and with older UTA line data in green.
\label{fig:u125col21n}}
\end{figure}

Finally, in Figure \ref{fig:u125col21n} we show the spectrum produced 
by an intervening cloud near the upper range of the cool stable branch.
This cloud has solar abundances, a column density of 
N(H) = 10$^{21}$\,cm$^{-2}$, and an ionization parameter
of log U = 1.25.
A portion of the coarse continuum which is used for
continuum radiative transfer and atomic rates is shown in the upper panel.
The assumed SED, which includes the ``Big Bump'' from the central accretion
disk and a non-thermal X-ray power law, has many absorption features 
superimposed.
Emission lines are also produced by the cloud but are weak in this portion of
the spectrum due to the low gas temperature.

The lower panel shows a small portion of the fine continuum in the
neighborhood of the $\sim 15$ \AA\ UTA feature.
Two curves are shown, with the solid line using the results presented in this paper, and the 
dashed line employing the other data sources summarized above.
Significant differences are present.

%%%%%%%%%%%%%%%  F11  %%%%%%%%%%%%%%%%%%%%%%%%%%%%%%%%%
%%%%%%%%%%%%%%%%%%%%%%%%%%%%%%%%%%%%%%%%%%%%%%%%

\section{Conclusions}

This paper has summarized advances in the atomic data needed to
simulate conditions in a photoionized plasma.
All of these improvements are now in the development version of the plasma
simulation code Cloudy, and will be part of the next major release.
The specific advances are the following:

\begin{itemize}

\item
We have calculated a large set of atomic data for UTA lines using the CI 
method implemented in the {\sc civ3} code. This new dataset substantially 
extends previous atomic data of \citet{fe1514} in two ways. 
First, the E1 transitions from the inner $2s$
shell are determined. Second, the data sets for \ion{Fe}{14}, \ion{Fe}{15} 
and \ion{Fe}{16} include lines $2l - 4l, 5l$ in addition to the earlier
determined lines $2p - 3l$.

\item
We have incorporated this large set of new UTA data into the spectral
simulation code Cloudy, and applied it to problems in AGN.
These improvements will be part of the next major release of the code.

\item
We summarize our data sources for UTA transitions.
There are still missing data, even for very important ions.
These should be a priority for new theoretical calculations.

\item
We summarize how our data, which were computationally very expensive to undertake,
compare to simpler calculations.
Line wavelengths differ (although insignificantly) due to differences in the 
computed level energies, but the transition rates are in good agreement.

\item
The UTA lines are often strongly damped, many having damping parameters
$a \gg 1$.  We have improved the form of the Voigt function used by
Cloudy to handle such strongly damped lines.

\item
We show how the \ion{Fe}{14} UTA at $\lambda15.5\,$\AA\ can be used
to measure the density of the gas, or identify whether the density 
is significantly above or below $10^9$ cm$^{-3}$.
Such measurements would help determine the location of the warm absorber
in AGN.

\item
The total ionization rate is increased by roughly 30\% with the new
set of UTA data, which have far more lines.
This changes the ionization of the gas and alters the thermal
properties of photoionized gas exposed to a typical AGN SED.

\item
We present a newly computed thermal stability ``S curve'' using the
new data.
We show that the Fe ions considered in this paper are produced in the
warmer parts of the cool thermally stable branch, an unstable region,
and in cooler parts of the middle stable branch.
As a result, these lines probe the portions of the S curve which
determine which cloud parameters can persist.
Future work will investigate the effects changes in these regions
have upon the observed spectrum.

\end{itemize}

\acknowledgments

GJF acknowledges support by NSF (0908877; 1108928; and 1109061), 
NASA (10-ATP10-0053, 10-ADAP10-0073, and NNX12AH73G), JPL (RSA No 1430426), 
and STScI (HST-AR-12125.01, GO-12560, and HST-GO-12309).
RK acknowledges support from the project VP1-3.1-{\v S}MM-07-K-02-013 
funded by European Social Fund under the Global Grant measure.
FPK is grateful to AWE Aldermaston for the award of a William Penney 
Fellowship. PvH acknowledges support from the Belgian Science Policy office
through the ESA PRODEX programme.

%\appendix
%\section{Appendix material}

\clearpage

\end{document}